\documentclass[fleqn,usenatbib]{mnras}

\usepackage{newtxtext,newtxmath}

\usepackage[T1]{fontenc}

\DeclareRobustCommand{\VAN}[3]{#2}
\let\VANthebibliography\thebibliography
\def\thebibliography{\DeclareRobustCommand{\VAN}[3]{##3}\VANthebibliography}

\usepackage{graphicx}	
\usepackage{amsmath}	

\def\commenta{$^*$}
\def\commentb{$^\dagger$}
\def\commentc{$^\ddagger$}
\def\commentd{$^\S$}
\def\commente{$^\|$}

\title[WZ Sge-type dwarf nova outbursts in TESS]{TESS and ground-based observations of WZ Sge-type dwarf novae in outburst}

\author[Y. Tampo et al.]{
Y. Tampo,$^{1,2}$\thanks{E-mail: yusuke@saao.ac.za}
N. Kojiguchi,$^{3,4}$
K. Isogai,$^{3,5}$
D. Nogami,$^{4}$
H. Itoh,$^{6}$
F.-J. Hambsch,$^{7,8,9}$
K. Matsumoto,$^{10}$
\newauthor
R. Matsumura,$^{10}$
D. Fujii,$^{10}$
T. Tordai,$^{11}$
Y. Sano,$^{12,13,14}$
B. Monard,$^{15}$
P.A. Dubovsky,$^{16, 17}$
T. Medulka,$^{16}$
\newauthor
D.A.H. Buckley,$^{1,2,18}$
N. Rawat,$^{1}$
S.B. Potter,$^{1,2,19}$
A. van Dyk,$^{1,2}$,
P.J. Groot,$^{1,2,20}$,
P. Woudt,$^{2}$
S. Kiyota,$^{21}$
\newauthor
G. Bolt,$^{22}$
T. Vanmunster,$^{9,23,24}$
J. Pietz,$^{25}$
P. Starr,$^{26}$
S.Y. Shugarov,$^{27,28}$
K. Kasai,$^{29}$
K. Menzies,$^{30}$
\newauthor
S.M. Brincat,$^{31}$
E.P. Pavlenko,$^{32}$
A. Baklanov,$^{32}$
J. Ito,$^{4}$
and
T. Kato$^{4}$
\\
$^{1}$South African Astronomical Observatory, PO Box 9, Observatory, 7935, Cape Town, South Africa\\ 
$^{2}$Department of Astronomy, University of Cape Town, Private Bag X3, Rondebosch 7701, South Africa\\ 
$^{3}$Okayama Observatory, Kyoto University, 3037-5 Honjo, Kamogatacho, Asakuchi, Okayama 719-0232, Japan\\ 
$^{4}$Department of Astronomy, Kyoto University, Kitashirakawa-Oiwake-cho, Sakyo-ku, Kyoto, 606-8502, Japan\\ 
$^{5}$Department of Multi-Disciplinary Sciences, Graduate School of Arts and Sciences, The University of Tokyo, 3-8-1 Komaba, Meguro, Tokyo 153-8902, Japan\\
$^{6}$Variable Star Observers League in Japan (VSOLJ), 1001-105 Nishiterakata, Hachioji, Tokyo 192-0153, Japan\\ 
$^{7}$Groupe Européen d’Observations Stellaires (GEOS), 23 Parc de Levesville, 28300 Bailleau l’Evêque, France\\ 
$^{8}$Bundesdeutsche Arbeitsgemeinschaft für Veränderliche Sterne (BAV), Munsterdamm 90, 12169 Berlin, Germany\\ 
$^{9}$Vereniging Voor Sterrenkunde (VVS), Zeeweg 96, 8200 Brugge, Belgium\\ 
$^{10}$Osaka Kyoiku University, 4-698-1 Asahigaoka, Osaka 582-8582, Japan\\ 
$^{11}$Polaris Observatory, Hungarian Astronomical Association, Laborc utca 2/c, 1037 Budapest, Hungary \\ 
$^{12}$Variable Star Observers League in Japan (VSOLJ), Nishi juni-jou minami 3-1-5, Nayoro, Hokkaido, Japan \\ 
$^{13}$Observation and Data Center for Cosmosciences, Faculty of Science, Hokkaido University, Kita-ku, Sapporo, Hokkaido 060-0810, Japan\\ 
$^{14}$Nayoro Observatory, 157-1 Nisshin, Nayoro, Hokkaido 096-0066, Japan \\ 
$^{15}$Kleinkaroo Observatory, Center for Backyard Astrophysics Kleinkaroo, Sint Helena 1B, PO Box 281, Calitzdorp 6660, South Africa\\ 
$^{16}$Vihorlat Observatory, Mierova 4, 06601 Humenne, Slovakia\\ 
$^{17}$Variable Star Section, Slovak Astronomical Society and Slovak Union of Astronomers, Slovakia\\ 
$^{18}$Department of Physics, University of the Free State, PO Box 339, Bloemfontein 9300, South Africa\\ 
$^{19}$Department of Physics, University of Johannesburg, PO Box 524, Auckland Park 2006, South Africa\\ 
$^{20}$Department of Astrophysics/IMAPP, Radboud University, PO Box 9010, 6500 GL Nijmegen, The Netherlands \\ 
$^{21}$Variable Star Observers League in Japan (VSOLJ), 7-1 Kitahatsutomi, Kamagaya, Chiba 273-0126, Japan\\ 
$^{22}$Camberwarra Drive, Craigie, Western Australia 6025, Australia\\ 
$^{23}$Center for Backyard Astrophycis Belgium, Walhostraat 1a, B-3401 Landen, Belgium\\ 
$^{24}$Center for Backyard Astrophycis Extremadura, e-EyE Astronomical Complex, ES-06340 Fregenal de la Sierra, Spain\\ 
$^{25}$Nollenweg 6, 65510 Idstein, Germany\\ 
$^{26}$Dubbo Observatory, 17L Camp Rd, Dubbo NSW 3830, Australia\\ 
$^{27}$Astronomical Institute of the Slovak Academy of Sciences, 05960 Tatranská Lomnica, Slovakia\\ 
$^{28}$Sternberg Astronomical Institute, Lomonosov Moscow State University, Universitetsky Ave.,13, Moscow 119992, Russia\\ 
$^{29}$Baselstrasse 133D, CH-4132 Muttenz, Switzerland\\ 
$^{30}$Center for Backyard Astrophysics (Framingham), MA 01701, USA\\ 
$^{31}$Flarestar Observatory, San Gwann SGN 3160, Malta\\ 
$^{32}$Federal State Budget Scientific Institution Crimean Astrophysical Observatory of RAS, Nauchny, 298409, Republic of Crimea\\ 
}

\date{Accepted XXX. Received YYY; in original form ZZZ}

\pubyear{\the\year{}}

\begin{document}
\label{firstpage}
\pagerange{\pageref{firstpage}--\pageref{lastpage}}
\maketitle

\clearpage
\begin{abstract}
Dwarf nova (DN) superoutbursts are accompanied by superhumps, which change their periods and profiles over a superoutburst. We present the TESS and ground-based observations of nine WZ Sge-type DNe and candidates in superoutburst. In TCP J23580961$+$5502508, ASASSN-23ba, PNV J19030433$-$3102187, V748 Hya, and ASASSN-25ci, we confirmed double-peaked oscillations called early superhumps, which are regarded as the unambiguous feature of WZ Sge-type DNe. On the other hand, the superhump and outburst properties of MO Psc and V1676 Her suggest that they may not be a member of WZ Sge-type DNe. The 2022 superoutburst of a confirmed WZ Sge-type DN TCP J05515391$+$6504346, however, lacked an early superhump phase. We find superhumps in a WZ Sge-type DN ASASSN-20mq during its rebrightening outburst.
Thanks to the continuous coverage of TESS, we find the broken-powerlaw rise of the outburst light curve in V748 Hya and PNV J19030433$-$3102187, previously found in only one WZ Sge-type DN observed by Kepler. Early superhumps appeared when the system reached $\simeq40$\% of the outburst peak flux. No orbital modulation from a hot spot is detected before and after this. This non-detection of orbital humps on the early rise of V748 Hya constrains that the corresponding mass transfer rate should be below $\simeq1\times10^{16}$ g s$^{-1}$, disfavouring an enhancement of a mass transfer rate by an order of magnitude or larger, even if it occurs. 
The contentious TESS observations also confirm the coexistence of early and ordinary superhumps during their transition and $\leq$2-cycle duration of stage A--B superhump transition in V748 Hya. 
\end{abstract}




\section{Introduction} \label{sec:intro}

Cataclysmic variables (CVs) are an interacting binary system hosting an accreting white dwarf (WD) as primary and a Roche-lobe-filling main-sequence star as secondary \citep[for a review, see][]{war95book, hel01book}. CVs are largely divided into magnetic and non-magnetic populations, based on whether accretion along the magnetic columns of the WD occurs or not. Dwarf novae (DNe) form a major subclass in non-magnetic CVs, which is characterized by quasi-periodic outbursts with an amplitude of 1--9 mag, duration of a few days--a month, and outburst cycle of weeks--decades. DN outbursts are understood in the thermal instability model \citep[for a review, see][]{osa96review, kim20thesis, ham20CVreview}. In this model, a cool disk in quiescence with neutral hydrogen increases its mass by a constant mass transfer from the secondary star. Once the disk reaches a critical surface density, hydrogen ionization triggers a rapid increase in disk viscosity and disk accretion rate, resulting in a hot and bright disk in outburst. After a disk depletes its mass, a cooling wave propagates through a disk, and the disk returns to quiescence.

During more energetic thermal-instability outbursts in accreting binary systems with a mass ratio ($q = M_{\text{secondary}} / M_{\text{WD}}$ where $M_{\text{secondary}} ~\text{and}~ M_{\text{WD}}$ are the secondary and WD mass, respectively) lower than 0.3, a periodic variation called superhumps is observed. They are characterized by a $\leq$0.3 mag amplitude and a period close to the orbital period \citep[][]{vog74vwhyi, pat81wzsge, nii21a18ey}. DN outbursts accompanied by superhumps are called a superoutburst, and DNe showing superoutbursts are classified as a SU UMa-type DN. 
The superhump period ($P_\text{SH}$) and profile evolve throughout a superoutburst, and are categorized into some superhump stages \citep[][]{Pdot}.  Ordinary superhumps are characterized by a single-peaked profile and a period slightly (1--3 \%) longer than the orbital period. This is understood as the consequence of the precession of an eccentric disk distorted because of the 3:1 tidal resonance \citep[tidal instability;][]{whi88tidal, hir90SHexcess}. Ordinary superhumps are further divided into three stages based on their period evolution: stage-A superhumps with growing amplitude and the longest and stable period, stage-B superhumps with the increasing period and decreasing amplitude, and stage-C superhumps with the shortest period.  Period derivatives of superhump period ($P_\text{dot} = \dot{P}_\text{SH} / P_\text{SH}$) are usually defined for stage B superhumps. In the thermal-tidal instability (TTI) model, stage-A superhumps are understood as the growth of tidal excitation at the 3:1 resonance radius, and stage-B superhumps correspond to the inward propagation of eccentricity across a disk \citep[][]{kat13qfromstageA}. 
A system with a mass ratio typically below 0.1 also shows early superhumps, double-peaked variations with a period almost equal \citep[$\simeq$0.1\% level;][]{pat81wzsge, ish02wzsgeletter} to the orbital period. These systems form a WZ Sge-type DN subclass \citep[see][for a review]{kat15wzsge}. The origin of early superhumps is attributed to the geometric effect of the vertically extended double-arm structure in a disk excited by the 2:1 resonance \citep[][]{lin79lowqdisk, osa02wzsgehump, uem12ESHrecon}. Because the 2:1 resonance suppresses the growth of the 3:1 resonance \citep[][]{lub91SHa}, early superhumps are always observed before ordinary superhumps. This also naturally leads to the long ($\geq$1 week) waiting time before the appearance of ordinary superhumps in a low-inclination system, where early superhump amplitude is below the detection limit.

Numerous works have been carried out on superhumps with ground observations \citep[e.g.][and their following series]{lem93tleo, Pdot}. However, they often suffer from observation gaps and different data quality from different telescopes.  The unpredictable and infrequent superoutbursts also make it difficult to study the early rise of an outburst. Compared to this, the Kepler space telescope and Transiting Exoplanet Survey Satellite (TESS) provide continuous observations with high and uniform precision of photometry \citep[][]{Kepler, K2mission, TESS}. Many CVs, including SU UMa-type DNe and superoutbursts, have been studied with Kepler and TESS \citep[e.g., ][]{can10v344lyr, woo11v344lyr, osa13v1504cygKepler, cou19zcha, liu23tesssuuma, boy24k2bs5}.  \citet{ver24awdqpo} presented the TESS observations of some WZ Sge-type DNe in quiescence. Examples of WZ Sge-type DNe in outburst with Kepler and TESS observations are, however, still limited due to their longer outburst cycles \citep[KSN:BS-C11a in Kepler and ASASSN-24hd in TESS; ][]{rid19j165350, tam25asassn24hd}, despite their large fraction in volume-limited samples \citep[][]{pal20GaiaCVdensity}.

These Kepler and TESS observations of WZ Sge-type DNe in outburst potentially give a clue to some of the important questions about their nature. 
First is the origin of their long outburst cycle over a decade \citep[][]{ort80wzsge, kat15wzsge} and an energetic outburst \citep[total accreted mass 1--2 $\times 10^{24}$ g during a superoutburst;][]{sma93wzsge} with very low mass transfer rate \citep[$\dot{M}_\text{tr} \simeq 1\times 10^{15}$ g s$^{-1}$;][]{sma93wzsge, ama21ezlyn, neu23bwscl}. In the framework of the TTI model, \citet{sma93wzsge, osa95wzsge, how95TOAD} pointed out that the quiescence viscosity needs to be as low as $\alpha \leq 0.001$ to prevent an outburst triggered within a shorter outburst cycle. 
On the other hand, \citet{las95wzsge, ham97wzsgemodel} discussed that, if the truncation of the inner disk due to either the WD magnetosphere or the evaporation effect works to prevent an inside-out outburst, a system can have a long superoutburst cycle even with a quiescence viscosity similar to that of SU UMa-type DNe ($\alpha \sim 0.01$). 
They also claimed that, in such cases, an outburst must be triggered and maintained by the enhanced mass transfer by a factor of $\geq$100 from the secondary star because of a shortage of disk mass stored before an outburst and a quasi-stable disk in quiescence. \citet{war96wzsge, mey98wzsge, mat07wzsgepropeller} discussed hybrid models with the inner disk truncation and outbursts triggered under the constant mass transfer rate and various values of quiescence viscosity. In a case where an outburst is triggered and maintained by the mass transfer burst from the secondary star, a natural expectation is enhanced emission from the hot spot on the outburst rise and maximum. However, such a behaviour has never been observed in WZ Sge-type DNe photometrically \citep[][]{ish02wzsgeletter, kat15wzsge} and spectroscopically \citep[][]{tov22v455andspec}, although the samples are very limited due to their unpredictable outbursts and long outburst cycles. Although KSN:BS-C11a was observed from its early outburst rise, its orbital profiles were not resolved due to the long cadence ($\simeq$30 min) of Kepler \citep[][]{rid19j165350}. Moreover, they reported the broken-powerlaw light curve in the outburst rise, which is still not observed in other WZ Sge-type DNe.  The unbiased observations by TESS are expected to provide more samples on the earliest outburst rise with better time resolution.

Another important aspect is understanding superhump evolution. Especially the superhump stage transitions are linked to the response of a disk to tidal forces and pressure effects in the TTI model, and are also potentially correlated with changes in X-ray properties \citep[][]{neu18j1222gwlib, nii21a18ey}. More continuous observations with TESS than with ground-based observations are suitable to study the timescales and smoothness of stage transitions in superoutbursts. \citet{tam25asassn24hd} reported that the superhump evolution of ASASSN-24hd is essentially identical to SU UMa-type dwarf novae observed with Kepler and TESS, given its relatively large mass ratio for a WZ Sge-type DNe. Since these superhump properties greatly depend on the mass ratio of a system \citep[][]{pat05SH, Pdot}, expanding samples into a wider range of mass ratios and unique outbursts is beneficial for improving our understanding of the TTI model.

In this paper, we report the TESS and ground-based observations of nine WZ Sge-type DNe and candidates in superoutburst with various orbital periods and mass ratios. The remainder of this paper is structured as follows. We present an overview of our samples and details of observations and analyses in Section \ref{sec:obs}. The results of our samples are presented in Section \ref{sec:result}. We discuss their superhump and light curve properties obtained from the TESS observations in Section \ref{sec:discussion}. Section \ref{sec:summary} summarises the discoveries of this paper.

\section{Observation and Analysis} \label{sec:obs}

\subsection{Target selection} \label{sec:target}

\begin{table*}
	\centering
	\caption{WZ Sge-type dwarf novae and candidates analysed in this paper.}
	\label{tab:target}
	\begin{tabular}{rcccccc} 
		\hline
		  Name & VSX type & Sector & Quiescence & Distance & Observers and observatories \commenta\\
		   & &  & (mag) & (pc) & & \\
		\hline
            ASASSN-20mq & UGWZ: & 31 & $G =$ 20.52(1) & 988$^{+673}_{-382}$ & MLF, HaC \\
    		MO Psc & UGWZ & 42 & $g =$ 22.0(1) & --- & --- \\
            TCP J23580961$+$5502508 & UGWZ & 57 & $G =$ 20.61(1) & 835$^{+628}_{-344}$ & BSM, HaC, Ioh, Kis, KU, OKU, San, Trt, Vih\\
            TCP J05515391$+$6504346 & UGWZ & 59,60 & $G =$ 20.58(1) & --- & \commentb CRI, Vih, Ioh, Kai, KU, MZK, OKU, PIE, Shu, Trt, Van\\ 
    		ASASSN-23ba & UGWZ & 62 & $B \simeq 21.61$ & --- & HaC, Ioh, MLF, SPE\\
            PNV J19030433$-$3102187 & UGWZ & 67 & $G =$ 20.71(1) & --- & GBo, HaC, Ioh \\
    		V1676 Her & UGWZ & 78,79 & $g =$ 21.50(8) & --- & --- \\
            V748 Hya & UGWZ & 89 & $G =$ 19.27(1) & 281$^{+19}_{-20}$ & HaC, Ioh, SAAO\\
            ASASSN-25ci & UGWZ: & 93 & $B \simeq 21.04$ & --- & HaC, MLF\\
		\hline
            \multicolumn{7}{l}{\commenta BSM=S.M. Brincat, CRI=Crimean Ast. Obs., GBo=G. Bolt, HaC=F.-J. Hambsch, Ioh=H. Itoh, Kai=K. Kasai, Kis=S. Kiyota, }\\
            \multicolumn{7}{l}{KU=Kyoto Univ., MLF=B. Monard, MZK=K. Menzies, OKU=Osaka Kyoiku Univ., PIE=J. Pietz,  Trt=T. Tordai,  }\\
            \multicolumn{7}{l}{SAAO=South African Ast. Obs., San=Y. Sano, Shu=S.Y. Shugarov, SPE=P. Starr, Van=T. Vanmunster,  Vih=Vihorlat Obs.}\\
            \multicolumn{7}{l}{\commentb The observers of TCP J05515391$+$6504346 are for its 2019 superoutburst.}\\
	\end{tabular}
\end{table*}

We first cross-checked the fields of view and epochs of the available TESS Full Frame Image (FFI) up to Sector 93 (June 2025) with known WZ Sge-type DNe and candidates in the AAVSO Variable Star Index \citep[VSX; ][]{VSX}. Our visual inspection yields nine WZ Sge-type DNe and candidates in outburst coinciding with the TESS observations, in addition to ASASSN-24hd in \citet{tam25asassn24hd}. 
These nine systems are summarised in Table \ref{tab:target}. Among our samples, ASASSN-20mq and ASASSN-25ci are listed as a WZ Sge-type DN candidate (UGWZ:) in VSX. We note that there is no literature reporting early superhumps for PNV J19030433$-$3102187, MO Psc, and V1676 Her but listed as a confirmed WZ Sge-type DN in VSX (see Sections \ref{sec:mopsc} \ref{sec:j1903}, and \ref{sec:herv1676}). The quiescence magnitude is available for all nine systems, and hence their outburst amplitudes are known. A distance with Gaia EDR3 \citep[][]{Bai21GaiaEDR3distance} is available for ASASSN-20mq, TCP J23580961$+$5502508, and V748 Hya. Two outbursts in our samples were observed across two sectors (TCP J05515391$+$6504346 in Sectors 59 and 60 and V1676 Her in Sectors 78 and 79), while the outbursts in the rest of our targets were observed only in one sector. Because their quiescence brightness is fainter than the typical limiting magnitude of TESS around 18 mag, we did not find any associated sources in other sectors of TESS.

\subsection{TESS observations} \label{sec:tessobs}

To obtain the TESS light curve, we used the FFIs calibrated through the TESS Science Processing Operations Center (SPOC) pipeline. We downloaded the 90$\times$90 pixels around the coordinate of our target via the Mikulski Archive for Space Telescopes (MAST) portal website. The photometry was performed using the \texttt{TESSreduce}\footnote{\url{https://github.com/CheerfulUser/TESSreduce/tree/master}} package \citep{rid21tessreduce}. We manually selected the frames before or after the outburst in a respective sector to be used for creating the reference image. After subtracting the reference image and backgrounds by \texttt{TESSreduce}, a PSF photometry was performed around the target following the default setups of this package. For TCP J05515391$+$6504346 and V1676 Her, we analysed two sectors independently.
This results in the differential flux $F_{\text{TESS}}$ (e s$^{-1}$). We converted this flux to relative magnitude $M_\text{TESS}$ (mag) simply assuming $M_\text{TESS} = -2.5 \log(F_{\text{TESS}})$.

After extracting the light curve with the \texttt{TESSreduce}, the global trend of the outburst rise, plateau, and rapid decline was removed by subtracting a smoothed light curve obtained by locally weighted polynomial regression \citep[LOWESS; ][]{LOWESS}. The superhump maxima were determined following \citet{Pdot, kat13j1939v585lyrv516lyr}. $O-C$ diagram of superhump maxima is used to present the evolution of the superhump period. The phase dispersion minimisation \citep[PDM;][]{PDM} method was applied for a period determination of superhumps in this paper. The 1$\sigma$ error for the PDM analysis is determined following \citet{fer89error, Pdot2}.

\subsection{VSNET observations} \label{sec:vsnet}

To cover the observation gaps in TESS, we also analysed the ground-based time-series observations obtained through the VSNET collaboration \citep[][]{VSNET}. We also obtained the time-resolved photometries with the Mookodi on the Lesedi telescope \citep[][]{era24mookodi} at the Sutherland observatory of the South African Astronomical Observatory. These observers and observatories are summarised in Table \ref{tab:target}. We reduced these data independently in the same manner as the TESS data; detrending of light curves by LOWESS, PDM for period determination, and $O-C$ diagram to track the evolution of superhump periods.

\subsection{Other time-domain surveys} \label{sec:otherobs}

We have also downloaded the archival datasets of other time-domain surveys, namely the All-Sky Automated Survey for SuperNovae (ASAS-SN) Sky Patrol \citep{ASASSN, koc17ASASSNLC, ASASSNV2}, the Asteroid Terrestrial-impact Last Alert System \citep[ATLAS; ][]{ATLAS, hei18atlas, smi20atlas, shi21atlas}, and the Zwicky Transient Facility \citep[ZTF;][]{ZTF, ztf_cite} broker Automatic Learning for the Rapid Classification of Events \citep[ALeRCE; ][]{alerce_ztf}, to study the overall profile of an outburst, rebrightening, and past activities. 
We did not use these survey data for the period analysis.

\section{Results} \label{sec:result}

\begin{figure}
	\includegraphics[width=0.95\linewidth]{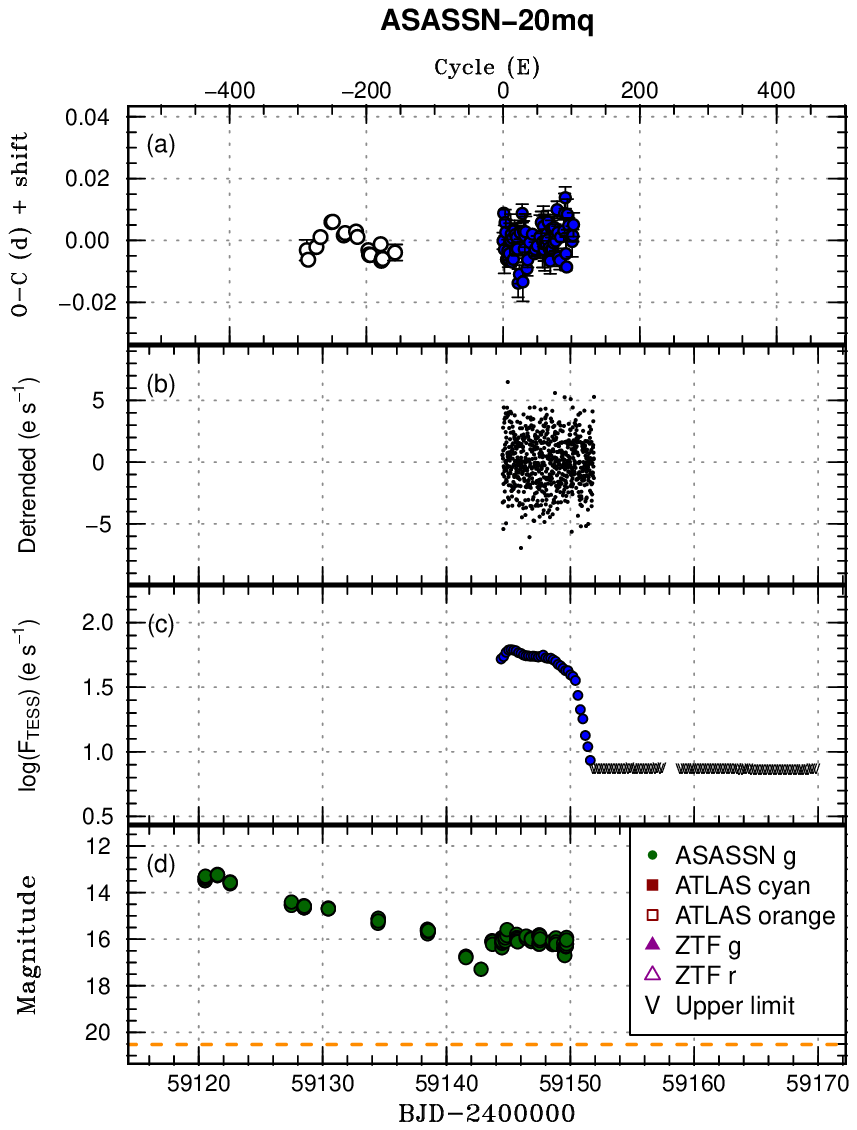}
    \caption{The outburst light curves and the $O-C$ diagram of superhump maxima of ASASSN-20mq.
                (a); The $O-C$ diagram of superhump maxima. $C = \text{BJD}~2459144.56543 + E\times0.05520$. The filled and open circles represent the superhump maxima determined using the TESS and VSNET observations, respectively.
                The VSNET data are vertically shifted ($-$0.025 d) for better visualisation.
                (b); detrended TESS light curve in the TESS flux scale (e s$^{-1}$). The light curve is binned in 0.01 d.
                (c); TESS light curve in a log-flux scale. The light curve is binned in 0.2 d. The v-shaped markers represent the 3$\sigma$ non-detection in the differential flux.
                (d); Corresponding light curve from other survey facilities. The filled circle, filled square, open square, filled triangle, and open triangle represent the data of ASAS-SN $g$ band, ATLAS $c$ band, ATLAS $o$ band, ZTF $g$ band, and ZTF $r$ band, respectively. The V-shaped markers represent the upper-limit observations from ATLAS. The horizontal dashed line shows the brightness of the quiescence counterpart in Table \ref{tab:target}.
            }
    \label{fig:oc-N1}
\end{figure}

\begin{figure*}
	\includegraphics[width=0.45\linewidth]{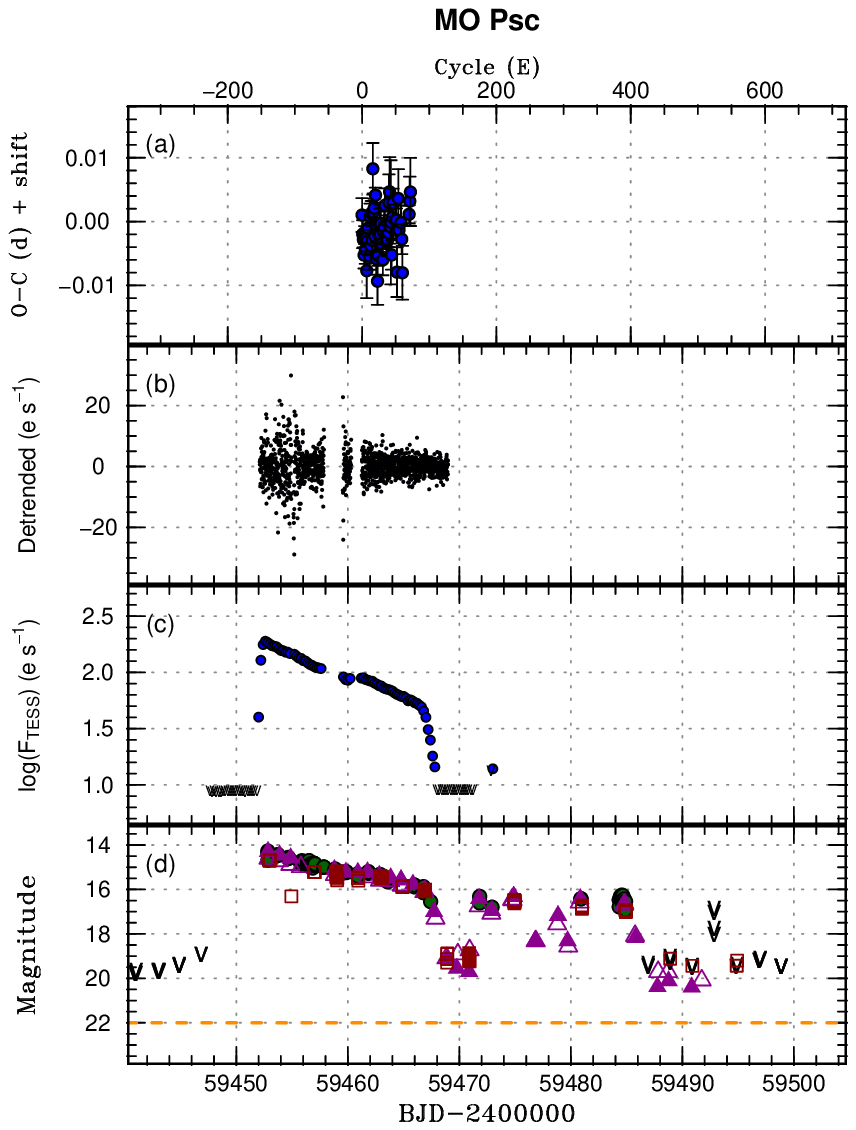}
	\includegraphics[width=0.45\linewidth]{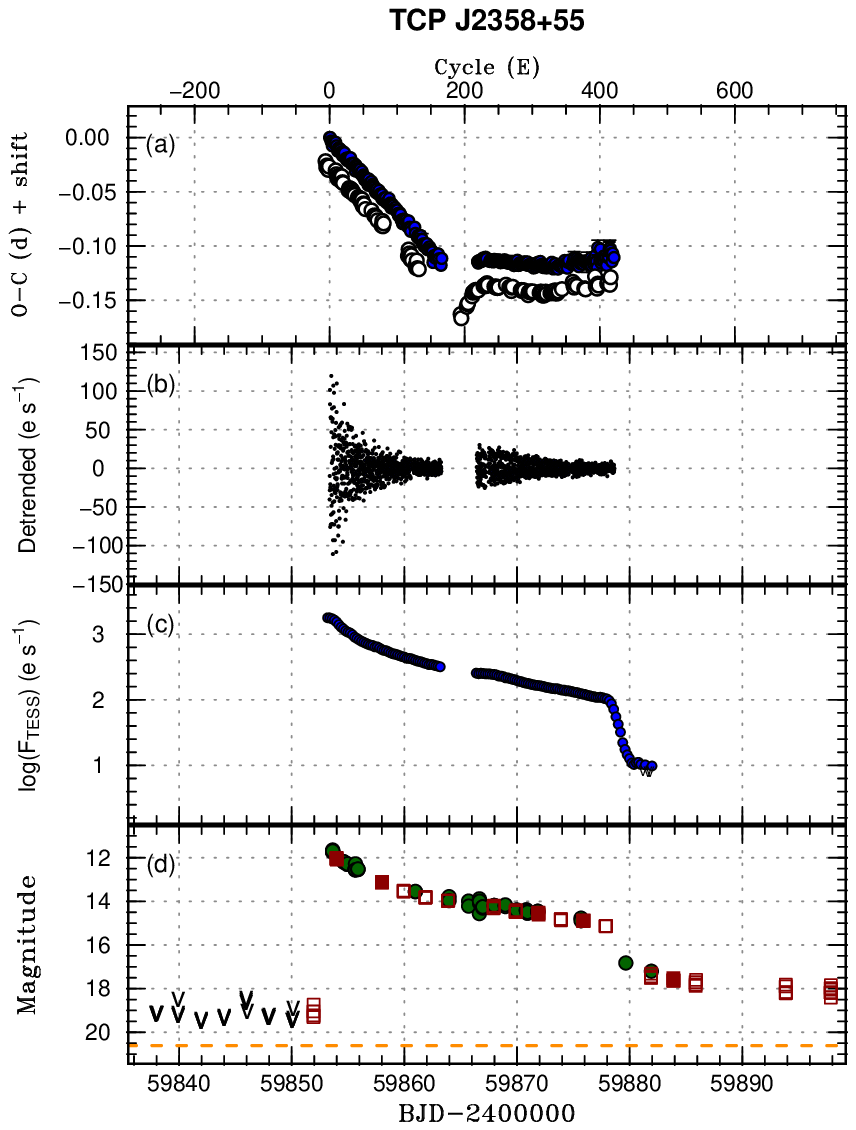}
    \caption{The same figure as Fig. \ref{fig:oc-N1} of MO Psc (left; {$C = \text{BJD}~2459461.27527 + E\times0.06022$.}) and TCP J2358$+$55 (right; {$C = \text{BJD}~2459853.39630 + E\times0.06006$.}).}
    \label{fig:oc-N2}
\end{figure*}

\begin{figure*}
	\includegraphics[width=0.45\linewidth]{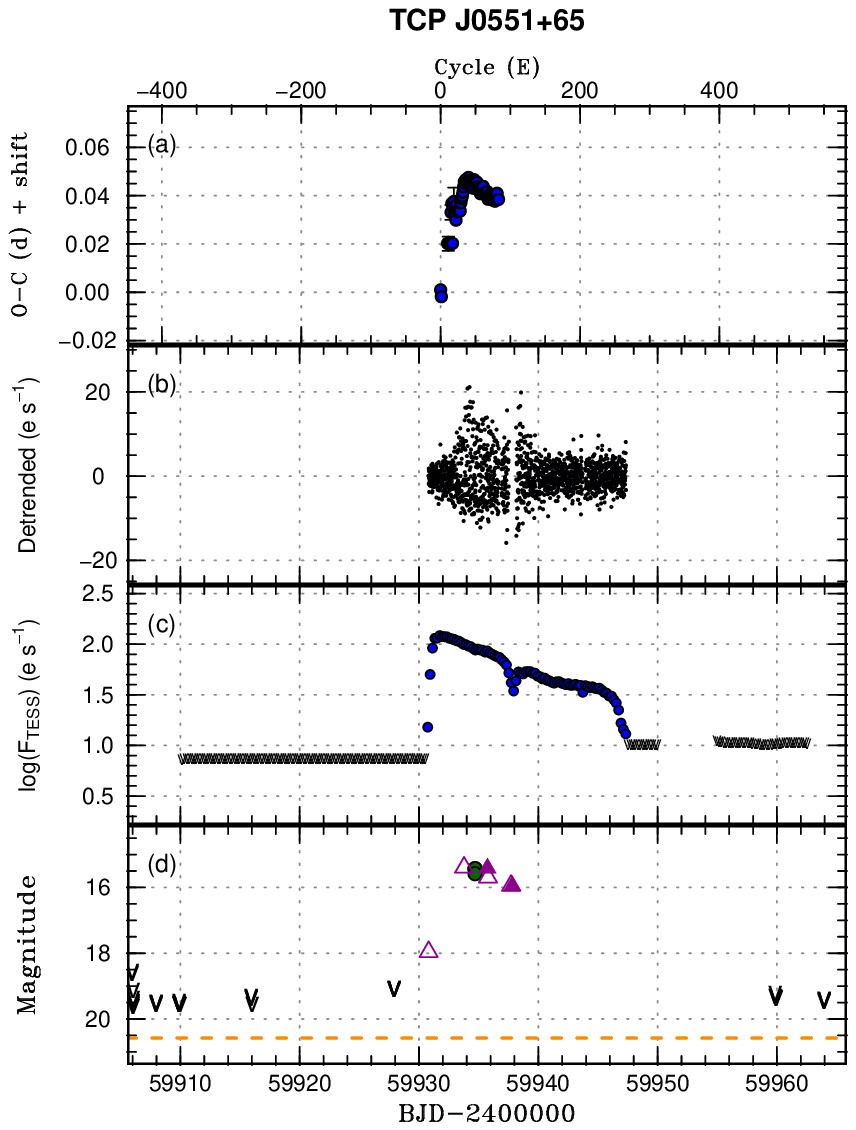}
	\includegraphics[width=0.45\linewidth]{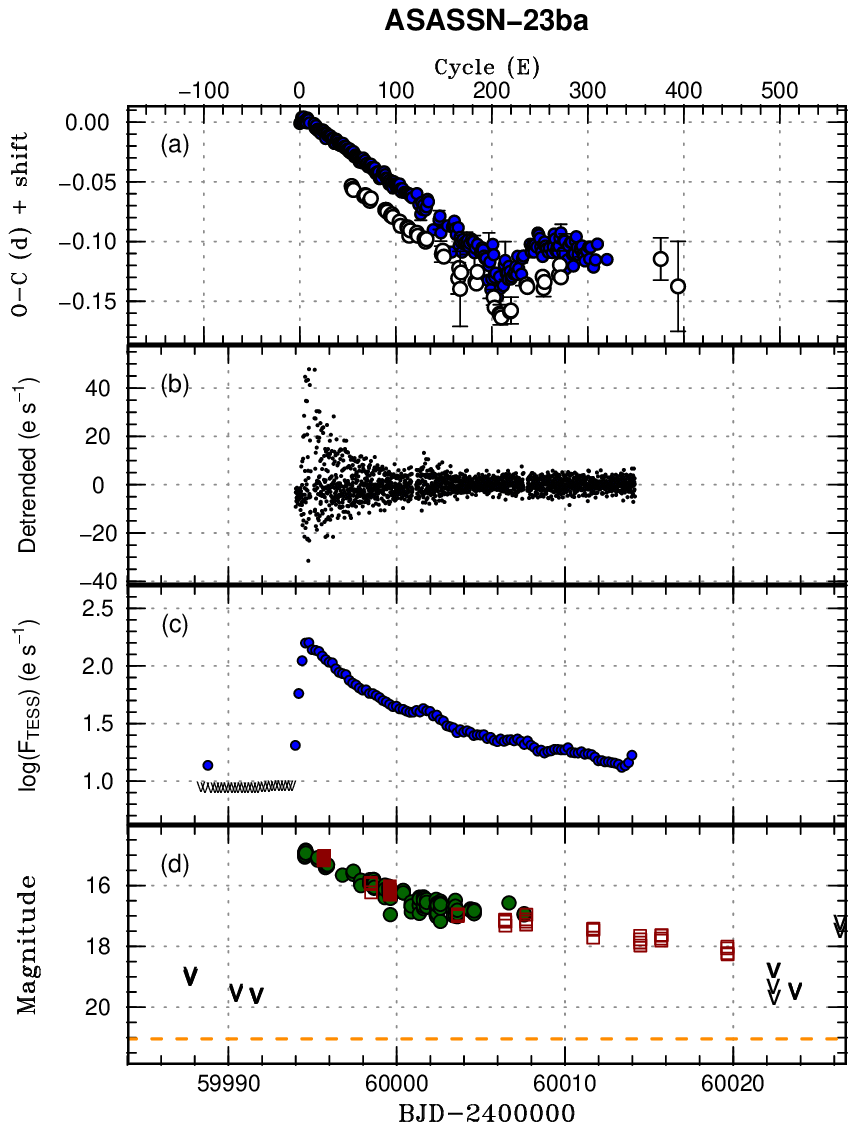}
    \caption{The same figure as Fig. \ref{fig:oc-N1} of TCP J0551$+$65 (left; {$C = \text{BJD}~2459931.79740 + E\times0.05838$.}) and ASASSN-23ba (right; {$C = \text{BJD}~2459994.26987 + E\times0.05720$.}). The dip in the middle of the outburst of TCP J0551$+$65 is due to the flaring backgrounds.}
    \label{fig:oc-N3}
\end{figure*}

\begin{figure*}
	\includegraphics[width=0.45\linewidth]{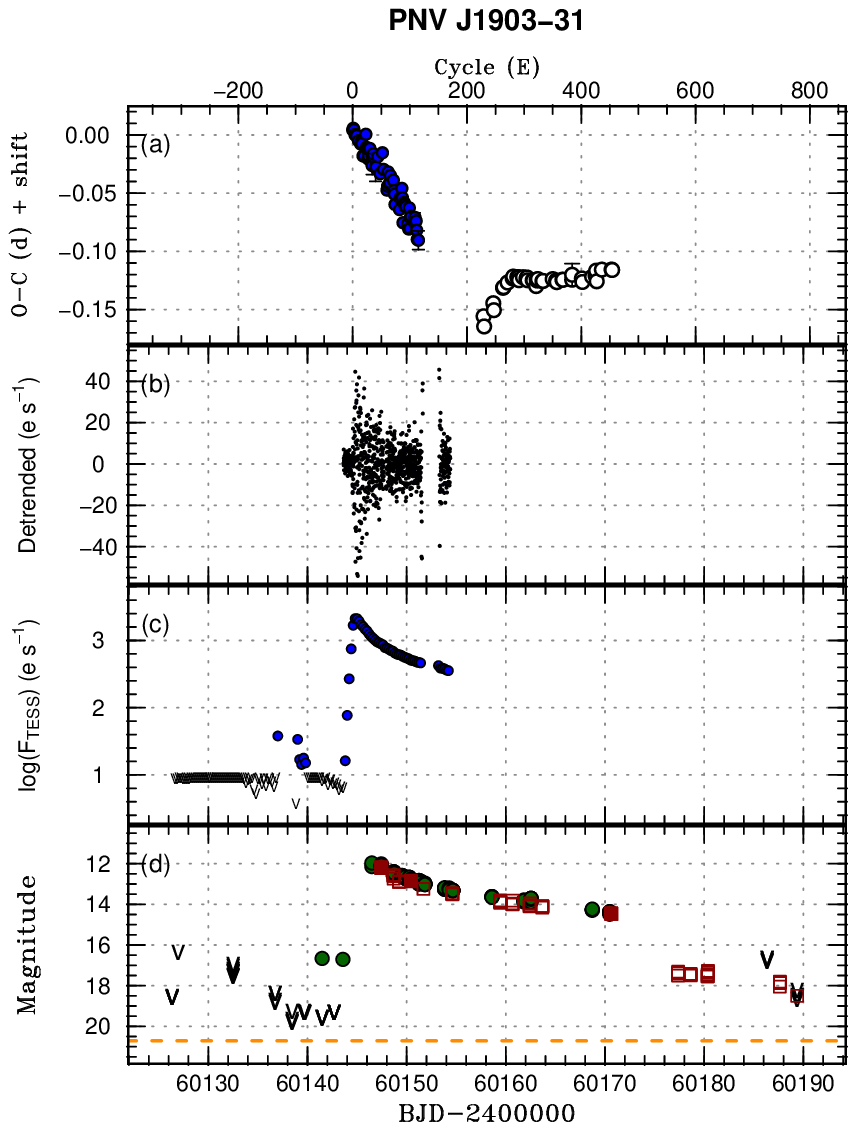}
	\includegraphics[width=0.45\linewidth]{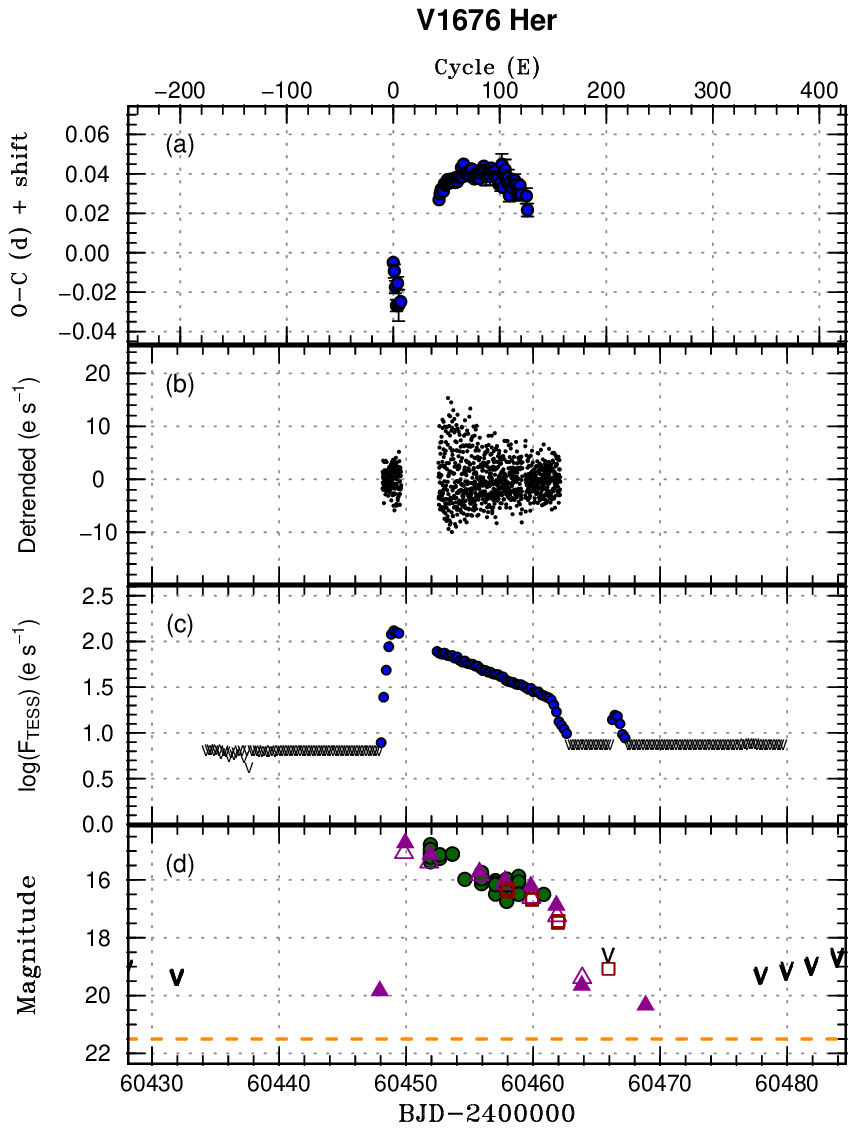}
    \caption{The same figure as Fig. \ref{fig:oc-N1} of PNV J1903$-$31 (left; {$C = \text{BJD}~2460144.58700 + E\times0.05769$.}) and V1676 Her (right; {$C = \text{BJD}~2460448.98684 + E\times0.08380$.}). The pre-outburst flares of PNV J1903$-$31 are due to the varying backgrounds and contaminations.}
    \label{fig:oc-N4}
\end{figure*}

\begin{figure*}
	\includegraphics[width=0.45\linewidth]{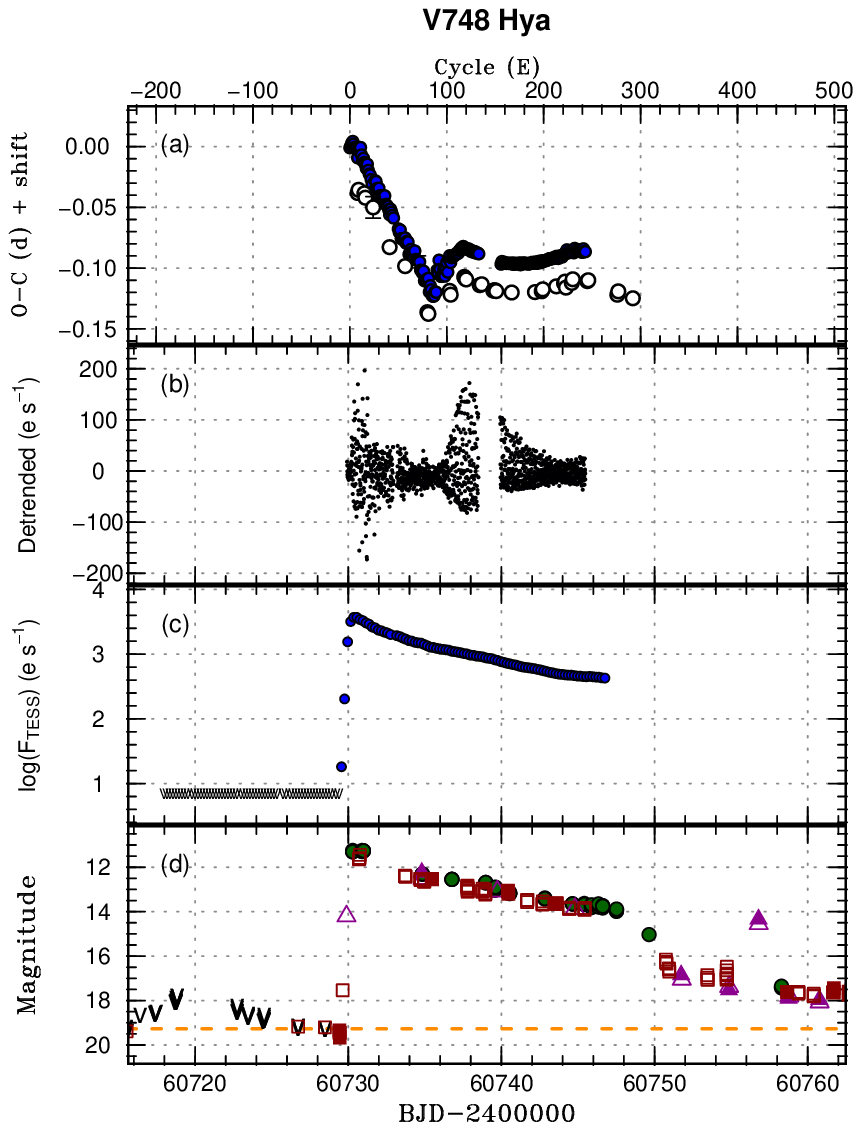}
	\includegraphics[width=0.45\linewidth]{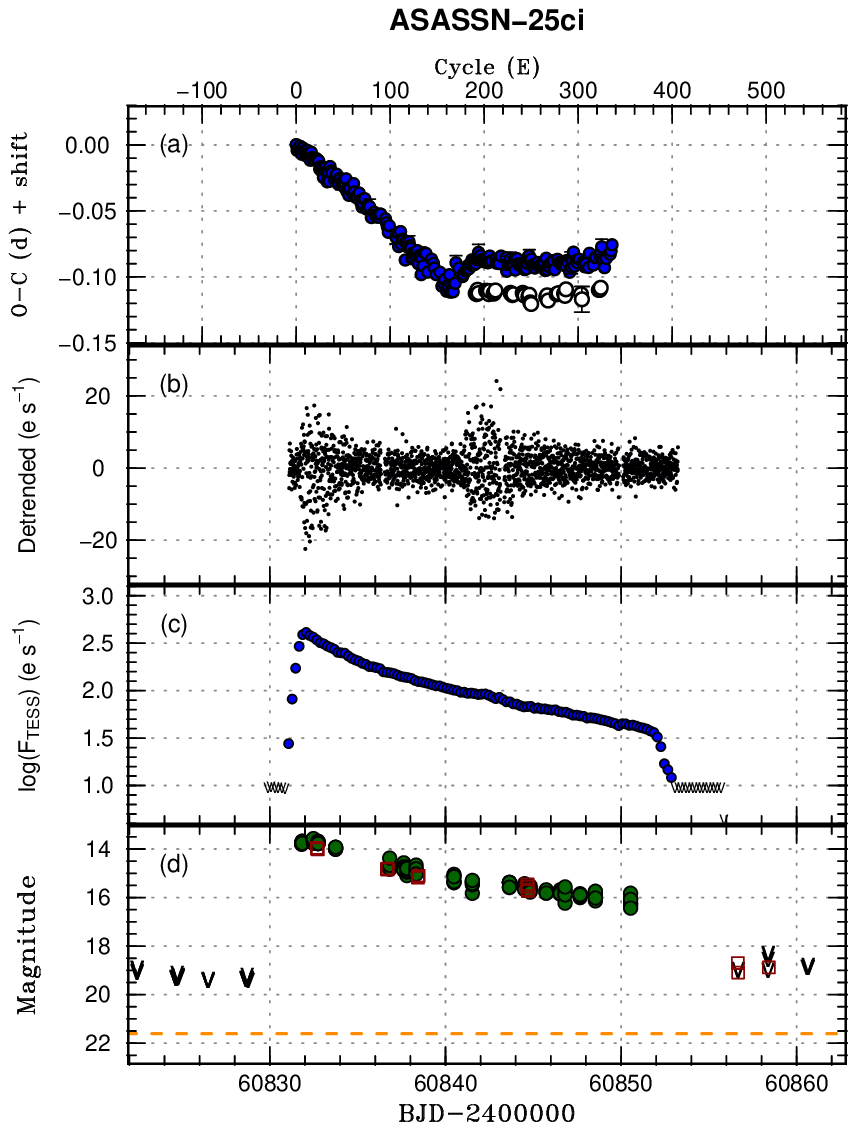}
    \caption{The same figure as Fig. \ref{fig:oc-N1} of V748 Hya (left; {$C = \text{BJD}~2460730.13400 + E\times0.063385$.}) and ASASSN-25ci (right; {$C = \text{BJD}~2460831.51575 + E\times0.053630$.}).}
    \label{fig:oc-N5}
\end{figure*}

\begin{figure*}
	\includegraphics[width=\linewidth]{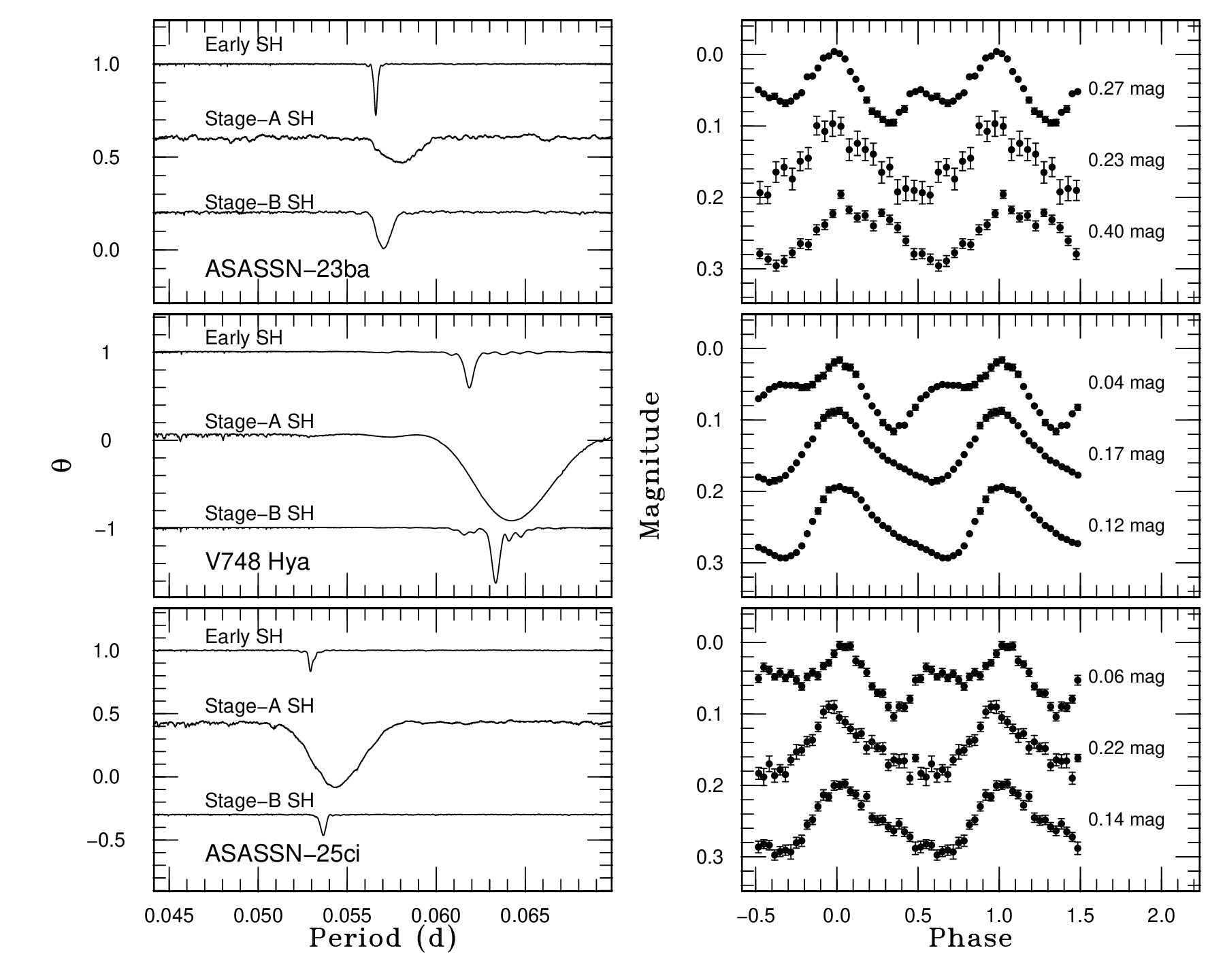}
    \caption{Results of PDM analysis (left) and phase-folded profiles (right) in selected time windows using the TESS observations.
                The superhump amplitudes are normalised to 0.1 mag for visualisation purposes. The actual amplitude is shown to the right in a magnitude scale. The superhump phase is arbitrarily determined.
                The top, middle, and bottom panels present the data of ASASSN-23ba (early, ordinary stage-A, and ordinary stage-B superhumps from top to bottom),  V748 Hya (early, ordinary stage-A, and ordinary stage-B superhumps from top to bottom), and ASASSN-25ci (early, ordinary stage-A, and ordinary stage-B superhumps from top to bottom).
            }
    \label{fig:pdm1}
\end{figure*}

\begin{figure*}
	\includegraphics[width=\linewidth]{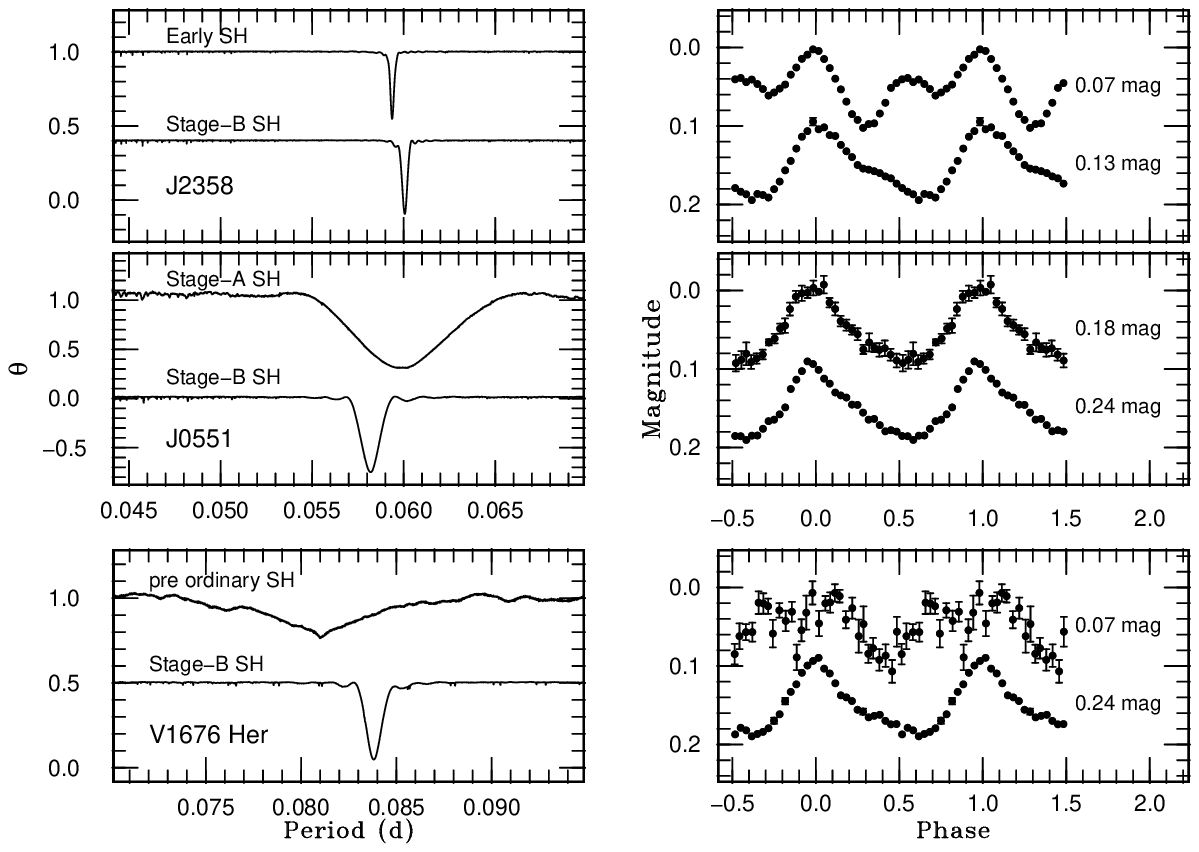}
    \caption{Same figure as Fig. \ref{fig:pdm1}, for 
                            TCP J2358$+$55 (early (top) and ordinary stage-B (bottom) superhumps), 
                            TCP J0551$+$65 (ordinary stage-A (top) and ordinary stage-B (bottom) superhumps), and
                            V1676 Her (pre-stage-A-superhump oscillation (top) and ordinary stage-B (bottom) superhumps)
                            from top to bottom.
            }
    \label{fig:pdm2}
\end{figure*}

\begin{figure*}
	\includegraphics[width=\linewidth]{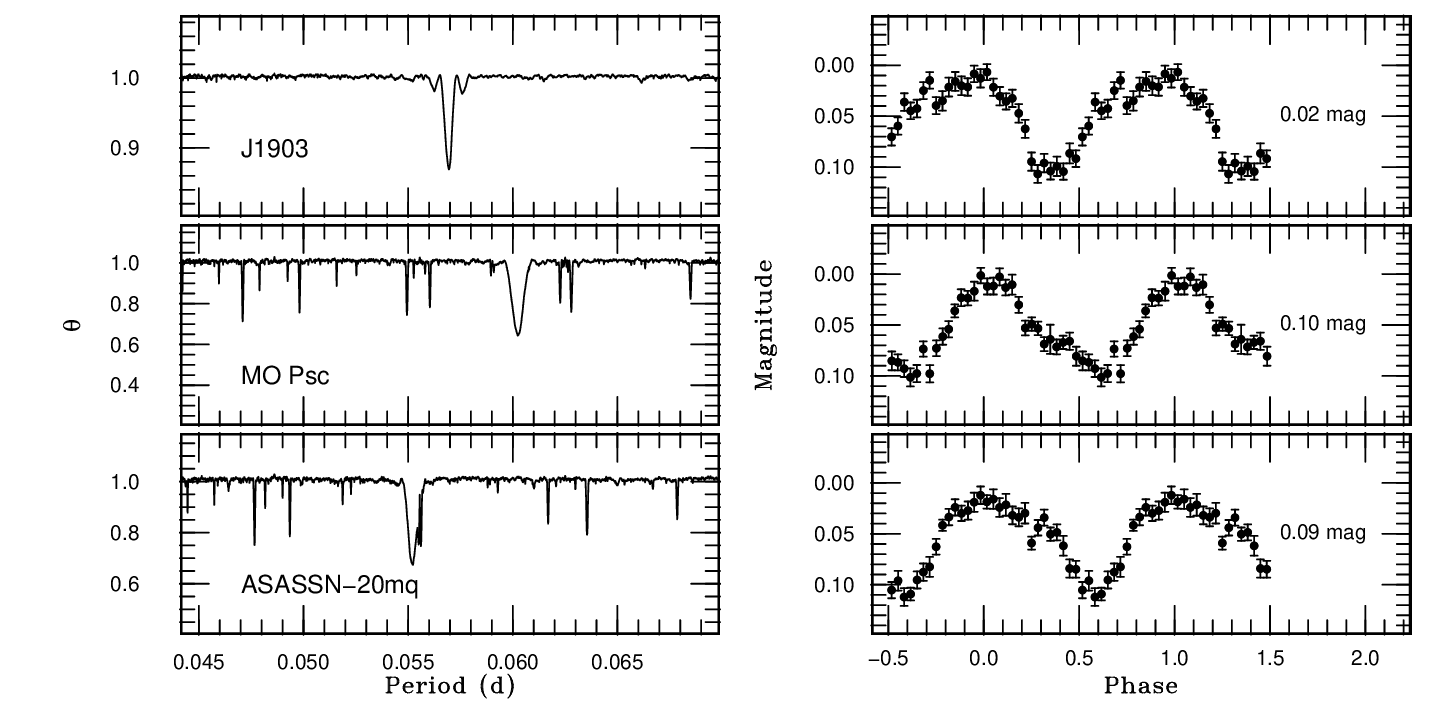}
    \caption{Same figure as Fig. \ref{fig:pdm1}, for 
                            PNV J1903$-$31 (early superhumps), 
                            MO Psc (superhumps during the outburst plateau), 
                            and ASASSN-20mq (superhumps during the rebrightening outburst)
                            from top to bottom.
            }
    \label{fig:pdm3}
\end{figure*}

\begin{figure}
	\includegraphics[width=0.95\linewidth]{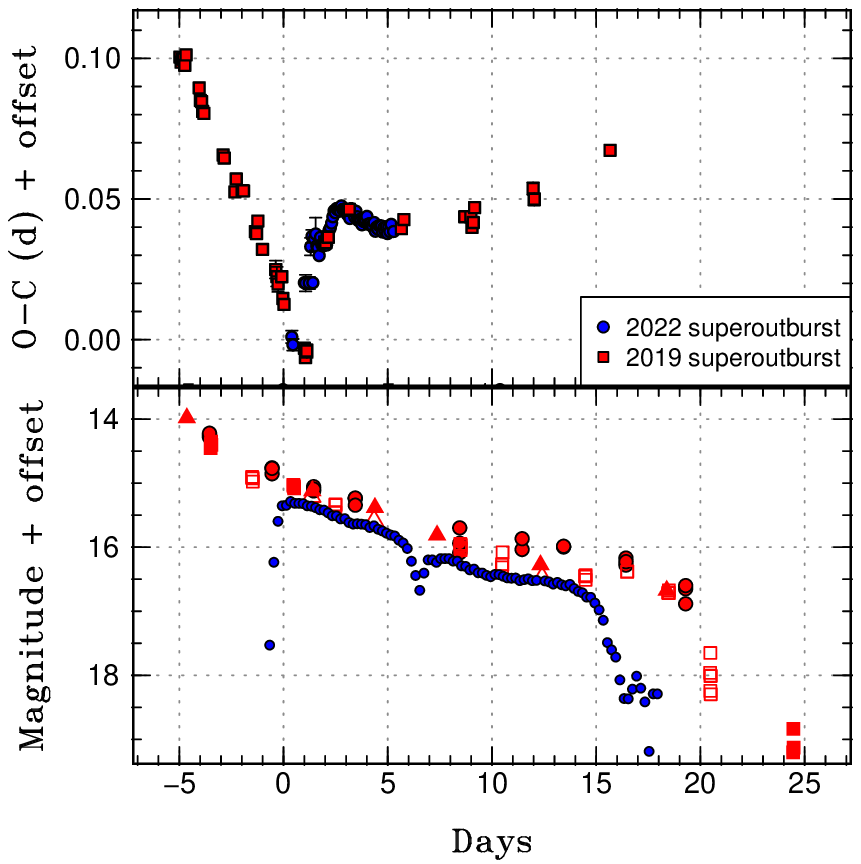}
    \caption{Comparison of the $O-C$ diagrams (top) and light curve (bottom) during the 2019 (red) and 2022 (blue) superoutbursts of TCP J0551$+$65. The symbols in the bottom panel are the same as Fig. \ref{fig:oc-N1}. The x and y axes are shifted to match the $O-C$ diagram around the stage A--B superhump transition.}
    \label{fig:ocj0551}
\end{figure}

\begin{table*}
	\centering
	\caption{Superhump periods $P_{\text{SH}}$, superhump period derivatives $P_{\text{dot}}$, and mass ratios $q$ of our targets based on the TESS observations.}
	\label{tab:superhumps}
	\begin{tabular}{rcccccc} 
		\hline
		  Name & $P_{\text{early~SH}}$ & $P_{\text{stage-A~SH}}$ & $P_{\text{stage-B~SH}}$ & $P_{\text{SH}}$ & $P_{\text{dot}}$ & $q$\\
		   & (d) & (d) & (d) & (d) & (cycle$^{-1}$) & \\
		\hline
            ASASSN-20mq & --- & 0.0559(1)\commenta & 0.05508(2)\commenta & 0.05520(1)\commentc & 7(4) $\times 10^{-5}$\commenta& --- \\
		      MO Psc & --- & --- & --- & 0.06025(2)\commentd & --- & --- \\
		      TCP J23580961$+$5502508 & 0.059371(3) & 0.06105(2)\commenta & 0.060059(3) & --- & 3.1(3) $\times 10^{-5}$ & 0.074(1) \\
            TCP J05515391$+$6504346 & 0.057345(8)\commenta & 0.0598(2) & 0.05819(1) & --- & 10(4) $\times 10^{-5}$ & 0.12(1) \\
		      ASASSN-23ba & 0.056605(4) & 0.05809(9) & 0.05705(3) & --- & 1(12) $\times 10^{-5}$ & 0.068(5) \\
		      PNV J19030433$-$3102187 & 0.05694(1) & 0.05863(5)\commenta & 0.057701(4)\commenta & --- & 2.9(6) $\times 10^{-5}$\commenta & 0.078(3) \\
    		V1676 Her & --- & --- & 0.08381(1) & 0.08102(6)\commente & $-$16(1)$\times 10^{-5}$ & --- \\
            V748 Hya & 0.061861(9) & 0.06423(5) & 0.063341(5) & --- & 12.5(2) $\times 10^{-5}$ & 0.103(3) \\
            ASASSN-25ci & 0.052948(8) & 0.05431(9) & 0.053656(9) & --- & 4.8(9) $\times 10^{-5}$ & 0.067(5) \\

		\hline
        \multicolumn{7}{l}{\commenta Based on the VSNET observations.}\\
        \multicolumn{7}{l}{\commentc Superhumps during the rebrightening outburst.}\\
        \multicolumn{7}{l}{\commentd Ordinary stage-B or stage-C superhumps.}\\
        \multicolumn{7}{l}{\commente Variations before the appearance of ordinary superhumps.}\\
	\end{tabular}
\end{table*}

Our results on individual outbursts are presented in Fig. \ref{fig:oc-N1}--\ref{fig:oc-N5}. The panel (d) presents the overall light curve around the outburst from ASAS-SN, ATLAS, and ZTF. The panels (c) and (b) show the raw and detrended light curves from the TESS observations. The panel (a) presents the $O-C$ diagram of superhump maxima based on the TESS (filled circles) and VSNET (open circles) observations. The $O-C$ of the VSNET observations is vertically shifted for visualisation purposes.
Table \ref{tab:superhumps} summarises the obtained superhump periods via the PDM analysis. The period derivative $P_{\text{dot}}$ is measured using the superhump maxima during the stage-B superhump phase. We also derived the mass ratio for the systems in the cases where both early and stage-A superhump periods are available, based on the relation between the superhump period excess during the stage-A superhumps and mass ratio \citep[][]{kat13qfromstageA, kat22updatedSHAmethod}. In all systems, we assume that the early superhump is equal to the orbital one, following the same manner as in the above papers. Fig. \ref{fig:pdm1} -- \ref{fig:pdm3} present the result of PDM analysis (left panels) and the phase-averaged profile with its best period (right panels).

\subsection{ASASSN-20mq} \label{sec:a20mq}

ASAS-SN discovered ASASSN-20mq in outburst at $g=$13.26 mag in September 2020.	Fig. \ref{fig:oc-N1} presents the obtained results of ASASSN-20mq.
The quiescence counterpart is Gaia EDR3 4791878210236290432 $G=20.52(1)$ mag at $988^{+673}_{-382}$ pc, hence the outburst amplitude is $\simeq$ 7.2 mag. 
The corresponding absolute magnitudes at outburst maximum and in quiescence are $M_{g}$ = 3.3(1.1) and $M_{G}$ = 10.5(1.1) mag, respectively, typical of WZ Sge-type DNe \citep[][]{tam20j2104}. 
The main outburst lasted for $\simeq$20 days, and was then followed by the rebrightening outburst for 6 d. Such a flat-top profile is known as the type-A rebrightening episode \citep[][]{ima06tss0222}.
ASASSN-20mq is suggested to be a WZ Sge-type DN based on this large outburst amplitude, the long waiting time before the appearance of ordinary superhumps, and its short superhump period of 0.05502(2) d (vsnet-alert 24781, 24792\footnote{All vsnet-alerts presented in this paper are available at \url{http://ooruri.kusastro.kyoto-u.ac.jp/mailarchive/vsnet-alert}}).

\subsubsection*{The 2020 superoutburst in TESS}

The TESS observation in Sector 31 covers only the rebrightening outburst (panels (c) and (d) of Fig. \ref{fig:oc-N1}). The PDM analysis during this rebrightening outburst yields a period of 0.05520(1) d. The phase-averaged profile shows a single and smooth peak with an amplitude of 0.09 mag (Fig. \ref{fig:pdm3}), typical of those observed during type-A rebrightenings \citep[e.g.][]{kim16alcom}.  We did not find any secure trends in the $O-C$ diagram during this rebrightening outburst. 

The VSNET observations detected ordinary superhumps during the main outburst. The periods of stage-A and stage-B superhumps are 0.0559(1) and 0.05508(2) d, respectively. This stage-B superhump period is shorter than that during the rebrightening outburst, consistent with other WZ Sge-type DNe \citep[][]{kat08wzsgelateSH}. The $P_{\text{dot}}$ during the stage-B superhumps is obtained as 7(4) $\times 10^{-5}$ cycle $^{-1}$. The ordinary superhumps appeared around BJD 2459128. Although the time-resolved observations before this date show small variations, no clear periodicity was found. We estimate the mass ratio of ASASSN-20mq as 0.09(3) based on the $P_{\text{dot}}$ and equation 30 in \citet{kat22updatedSHAmethod}. Hence, we confirm the previous classification of ASASSN-20mq as a WZ Sge-type DN with a low inclination because of this long ($\simeq$1 week) waiting time before the appearance of ordinary superhumps, its short superhump period, large outburst amplitude, long outburst duration, and lack of any previous outbursts.

\subsection{MO Psc} \label{sec:mopsc}

MO Psc (=USNO-B1.0 0915-0569037=OT J231110.9$+$013003) was first discovered in outburst by K. Itagaki in 2007. This system underwent another outburst in 2021, and \citet{shu21mopsc} suggested MO Psc as a WZ Sge-type DN candidate based on their claimed superhump period of 0.05161 d during the rapid decline of its 2021 superoutburst. The quiescence counterpart is Pan-STARRS release 1 (PS1) 109803477953651318 $g=22.0(1)$ mag \citep[][]{panstarrs1DR2}. There are no associated Gaia EDR3 objects. 
The 2021 superoutburst peaked on BJD 2459452 at around $g=$14.3 mag (left panels of Fig. \ref{fig:oc-N2}). The rapid decline from the outburst was observed on BJD 2459467. Hence, the outburst amplitude was 7.5 mag while the outburst duration was just 15 d, one of the shortest among known WZ Sge-type DNe \citep[e.g. see fig. 11 of ][]{tam24j0302}. Moreover, the outburst was followed by a complex series of rebrightening outbursts, likely type-B rebrightening series \citep[multiple rebrightening outbursts; ][]{ima06tss0222}.

\subsubsection*{The 2021 superoutburst in TESS}

TESS in sector 42 covers this 2021 superoutburst from the outburst rise to the start of the rebrightening outbursts. TESS records the outburst rise on BJD 2459452 (see Fig. \ref{fig:riselc} as well). The rise timescale is 0.125(3) d mag$^{-1}$.
We detect the superhumps on BJD 2459461 -- 2459466 with a period of 0.06025(2) d. The phase-averaged light curve shows a single-peaked profile with an amplitude of 0.10 mag (Fig. \ref{fig:pdm3}). There was no secure change of superhump periods, so we tentatively assign them as either ordinary stage-B or stage-C superhumps. We note that, due to the large scatter of the background, no significant period is detected during the first two quarters of the sector. 
We did not detect any significant periodicity during the rapid decline in the TESS data (BJD 2459467--2459468).

The obtained superhump period in TESS does not agree with the previously reported superhump period during the rapid decline phase. Since the TESS observations present this periodic variation for 40 cycles in the superoutburst plateau, the detected period is likely true. Although superhump period changes during an outburst, the difference of more than 15\% is too large compared to any reported changes \citep[e.g.][]{Pdot}. 
With this relatively long (0.06025(2) d) superhump period rather than the previously reported one \citep[0.05161 d;][]{shu21mopsc}, together with its short outburst duration ($\simeq$15 d), we suspects that MO Psc could be an SU UMa-type DN with no excitation of the 2:1 resonance, nevertheless the repeating rebrightening outbursts is commonly observed only in evolved WZ Sge-type DNe and SU UMa-type with long ($\geq 0.1$ d) superhump period \citep[][]{mro15OGLEDNe, kat20asassn14ho}. Outbursts with a similar duration were observed in a WZ Sge-type DN RZ Leo, although its orbital period is even longer \citep[0.07604 d;][]{ish01rzleo, Pdot} and it showed just a single rebrightening outburst. Another possibility is that this 2021 superoutburst lacked an early superhump phase despite MO Psc being a system with a mass ratio low enough to trigger the 2:1 resonance \citep[see Section \ref{sec:j0551} and e.g.][]{kim16alcom}. However, such an outburst usually follows within a few years after a more energetic superoutburst with early superhumps. Non-detection of any outbursts in ASAS-SN since October 2012 up to this 2021 superoutburst disfavours this scenario. Future studies of MO Psc either in quiescence or next outburst are required to confirm its WZ Sge-type nature.

\subsection{TCP J23580961$+$5502508} \label{sec:j2358}

TCP J23580961$+$5502508 (hereafter TCP J2358$+$55)\footnote{\url{http://www.cbat.eps.harvard.edu/unconf/followups/J23580961+5502508.html}} was discovered at 12.0 mag (unfiltered) by T. Kojima in outburst in September 2022, and peaked at $g$ = 11.7 mag on BJD 2459863.6 according to ASAS-SN (right panels of Fig. \ref{fig:oc-N2}). TCP J2358$+$55 was classified as a WZ Sge-type DN based on the detection of 0.2-mag-amplitude early superhumps (vsnet-alert 26954).
The counterpart is Gaia EDR3 1993876650919517440 $G=$20.61(1) mag, hence the outburst amplitude is $\simeq$ 8.6 mag. The corresponding distance is 835$^{+628}_{-344}$ pc, resulting in the absolute magnitudes at outburst maximum and in quiescence are $M_g$ = 2.1(1.2) and $M_G$ = 11.0(1.2) mag, respectively. The outburst duration was $\simeq$24.5 d from the outburst maximum to the start of the rapid fading (BJD 2459878.2). No rebrightenings were recorded in ATLAS and ASAS-SN.

\subsubsection*{The 2022 superoutburst in TESS}

TCP J2358$+$55 was observed by TESS in Sector 57. These TESS observations cover its 2022 superoutburst just after the outburst maximum to the end of the rapid decline phase.
We detect superhumps throughout the superoutburst. Based on the $O-C$ diagram and phase-averaged profile, the early and stage-B superhump phases correspond to BJD 2459852.5 -- 2459863.5 and BJD 2459867.75 -- 2459877.5, respectively. We determined their superhump periods as 0.059371(3) and 0.060059(3) d, respectively (Fig. \ref{fig:pdm2}). $P_{\text{dot}}$ in stage-B superhumps is 3.1(3) $\times 10^{-5}$ cycle $^{-1}$. The mean amplitude of early superhumps is 0.07 mag (Fig. \ref{fig:pdm2}). 
Judging from the $O-C$ diagram, the stage-A superhump is mostly missed due to the observation gap between the second and third quarters. Using the VSNET data, we obtained the stage-A superhump period as 0.06105(2) d. Together with its early superhump period, the mass ratio-superhump excess relation yields its mass ratio as 0.074(1).

\subsection{TCP J05515391$+$6504346} \label{sec:j0551}

TCP J05515391$+$6504346 (hereafter TCP J0551$+$65 \footnote{\url{http://www.cbat.eps.harvard.edu/unconf/followups/J05515391+6504346.html}}) was discovered in 2019 April and classified as a WZ Sge-type DN (vsnet-alert 23127). This 2019 superoutburst showed clear early superhumps with an amplitude of 0.16 mag and a period of 0.057345(8) d (vsnet-alert 23143; Fig. \ref{fig:ocj0551}). 
The quiescence counterpart is Gaia EDR3 287903786839962624 $G=20.58(1)$ mag. Although there was no record of rebrightening outbursts right after the 2019 superoutburst, ZTF (ZTF19aaofigg in ALeRCE\footnote{\url{https://alerce.online/object/ZTF19aaofigg}}) has detected further outbursts in July 2021 at $r\simeq17.5$ mag and in December 2022 (left panel of Fig. \ref{fig:oc-N3}), $\simeq$850 and 1350 days after the discovery outburst. This 2022 outburst peaked around $r=15.39$(1) mag on BJD 2459934, and entered the rapid decline on BJD 2459946.5, giving the outburst duration of $\simeq$12.5 d. These numbers are considerably smaller than those in its 2019 superoutburst, which peaked at 13.2 mag (unfiltered) and lasted for $\simeq$25 d (see Fig. \ref{fig:ocj0551}). The 2021 outburst appears even fainter (peak at $r\simeq17.5$) and shorter (5 detections only in one night), which can be a normal outburst or even a non-astrophysical event.

\subsubsection*{The 2022 superoutburst in TESS}

This 2022 superoutburst was covered in the TESS Sectors 59 and 60. The rise timescale is measured as 0.070(4) d mag$^{-1}$. Single-peaked variations are clearly visible at the outburst maximum on around BJD 2459932.4. These variations show a period of 0.0598(2) d, significantly longer than the early superhumps observed in its 2019 superoutburst. Based on the $O-C$ diagram, the stage A--B superhump transition occurred around BJD 2459934.1 or $E\simeq40$. We determined the periods of stage-A and stage-B superhumps as 0.0598(2) and 0.05819(1) d, respectively (Fig. \ref{fig:pdm2}). Considering this appearance of ordinary superhumps at outburst maximum, fainter outburst maximum, and shorter outburst duration, we conclude that this 2022 superoutburst of TCP J0551$+$65 lacked the early superhump phase. 
The large scatter of the backgrounds caused the non-real dip around BJD 2459938. We were not able to determine the individual superhump maxima in Sector 60 due to the lower counts. The PDM analysis using the data between BJD 2459940 and 2459946 yields a superhump period of 0.05840(3) d. This is slightly longer than the stage-B superhump period detected at the end of Sector 59, consistent with the later phase of stage-B superhumps. By combining its early superhump period in 2019 and stage-A superhump period in 2022, we obtain its mass ratio as 0.12(1). This is relatively large for being a WZ Sge-type DN.

Based on this fact, TCP J0551$+$65 is another example of a relatively rare population in WZ Sge-type DNe, which have shown superoutbursts both lacking and accompanying early superhump phase \citep[e.g., AL Com, V627 Peg, and V3101 Cyg; ][]{kim16alcom, tam20j2104, tam23v627peg}. Fig. \ref{fig:ocj0551} compares the $O-C$ diagram and light curves of the 2019 and 2022 superoutbursts. As mentioned above, the 2022 superoutburst significantly peaked fainter ($r\simeq$15.39(1) mag) and lasted shorter ($\simeq$12.5 d) than those of the 2019 superoutbursts (13.2 mag and $\simeq$25 d). Judging from the $O-C$ diagram, the duration of the stage-A superhump stage is similar to each other ($E\simeq$40), while that of stage-B superhumps is significantly longer in the 2019 superoutburst with early superhumps. The outburst interval between these two superoutbursts of TCP J0551$+$65 is 3.7 years, lying between AL Com (1.5 years) and V627 Peg (4.5--6.7 years), while V3101 Cyg showed three superoutbursts lacking early superhumps during its rebrightening series \citep[][]{tam20j2104}.

\subsection{ASASSN-23ba} \label{sec:a23ba}

ASAS-SN discovered ASASSN-23ba in February 2023, peaking at 14.6 mag on BJD 2459994.5 (right panels of Fig. \ref{fig:oc-N3}). The quiescence counterpart is GSC S681028508 $B\simeq$21.61 mag \citep[][]{GSC2}, hence the outburst amplitude is 7.0 mag. There are no associated Gaia EDR3 objects. Soon after the outburst detection, early superhumps with a period of 0.05662(1) d were detected (vsnet-alert 27444, 27458), confirming its WZ Sge-type nature. The superoutburst lasted until BJD 2460019.6; thus, the superoutburst duration is $\simeq$25 d. ASASSN-23ba did not show any rebrightening outbursts.

\subsubsection*{The 2023 superoutburst in TESS}

TESS Sector 62 observed ASASSN-23ba from the outburst rise to the middle of the superoutburst. The rise timescale is measured as 0.208(3) d mag$^{-1}$. The early superhumps with a mean amplitude of 0.27 mag were clearly detected (Fig. \ref{fig:pdm1}). Based on the $O-C$ diagram and superhump profiles, we determined the superhump stages as follows; early superhumps: BJD 2459994.3 -- 2460005.2, stage-A superhumps: BJD 2460006.1 -- 2460007.7, and stage-B superhumps: BJD 2460008.3 -- 2460014.1. The early, ordinary stage-A and stage-B superhump periods are determined as 0.056605(4), 0.05809(9), and 0.05705(3) d, respectively. $P_\text{dot}$ in stage-B superhump is poorly constrained as 1(12) $\times 10^{-5}$ due to the low SNR during the ordinary superhumps. Based on these periods, its mass ratio is obtained as 0.068(5).

\subsection{PNV J19030433$-$3102187} \label{sec:j1903}

PNV J19030433$-$3102187 (hereafter PNV J1903$-$31\footnote{\url{http://www.cbat.eps.harvard.edu/unconf/followups/J19030433-3102187.html}}) was discovered at 11.4 mag (unfiltered) by the New Milky Way team in July 2023 (left panels of Fig. \ref{fig:oc-N4}). The subsequent time-resolved observations detected ordinary superhumps, with a period of 0.05777(2) d, 15 d after the discovery, suggesting the WZ Sge-type DN classification (vsnet-alert 27932).
The counterpart is Gaia EDR3 6760477849042929664 $G=20.71(1)$ mag, and thus the outburst amplitude is $\approx 9.3$ mag. There is no parallax available in Gaia EDR3. The outburst continued from BJD 2460144 to at least BJD 2460170.76, for $\geq$26 d. No rebrightenings were recorded, although there are some observation gaps around the rapid decline, and it cannot be concluded. 
We note that the pre-outburst detections in ASASSN are contaminated by the nearby star, and the varying backgrounds in TESS, given the continuous non-detections in ATLAS at $\simeq19$ mag.

\subsubsection*{The 2023 superoutburst in TESS}

TESS Sector 67 covered PNV J1903$-$31 from the pre-outburst quiescence to the first $\simeq$10 days of the outburst. The rise timescale is 0.169(1) d mag$^{-1}$. The PDM analysis in outburst using the TESS data yields variations with a period of 0.05694(1) d (Fig. \ref{fig:pdm3}). Considering its flat-top profile, small amplitude ($\simeq$0.02 mag), no clear evolution of superhump period, and shorter period than that of ordinary superhumps detected through the VSNET observations, this is most likely early superhumps.

We have also analysed the VSNET observations during the ordinary superhump phase. These establish its stage-A and stage-B superhump periods as 0.05863(5) and 0.057701(4) d, respectively. This implies its mass ratio of 0.078(3).

\subsection{V1676 Her} \label{sec:herv1676}

V1676 Her \citep[= MASTER OT J165236.22+460513.2;][]{den13j1652atel4881} was discovered in March 2013 at 14.8 mag. The quiescence counterpart is PS1 163302531511155564 $g=21.50(8)$ mag \citep[][]{panstarrs1DR2}. There are no associated Gaia EDR3 objects. \citet{Pdot5} suggested that V1676 Her might be a period bouncer system in WZ Sge-type DNe, based on its large outburst amplitude ($\geq$7 mag) and long superhump period ($P_\text{stage-B~SH}$ = 0.08473(9) d) despite its red colour of the PS1 counterpart.
We find that V1676 Her went into another superoutburst in May 2024, which is also observed by TESS (right panels of Fig. \ref{fig:oc-N4}). This superoutburst reached the outburst maximum at $g=14.718(4)$ mag on BJD 2460449.9. Provided the start of the rapid decline on BJD 2460461.6, the superoutburst duration is $\simeq$12.5 d. A single and short rebrightening outburst was observed on BJD 2460466.5. These numbers are comparable to those of its 2013 superoutburst. In addition, there was no precursor outburst.

\subsubsection*{The 2024 superoutburst in TESS}

The 2024 superoutburst of V1676 Her was observed in the TESS Sectors 78 and 79. We detected the ordinary superhumps with a period of 0.08381(1) d and an amplitude of 0.24 mag in Sector 79 (i.e., after BJD 2460452.5; Fig. \ref{fig:pdm2}). This period roughly agrees with the superhump period reported in \citet{Pdot5}. Given the varying superhump period, this is likely the stage-B superhumps. The $P_\text{dot}$ during this stage is $-$16(1)$\times 10^{-5}$. Such a negative and large absolute value of $P_\text{dot}$ is typical for long-superhump-period and large-mass-ratio systems in \citet{Pdot}. The large superhump amplitude also implies a large mass ratio. 
Moreover, the PDM analysis at outburst maximum (BJD 2460448.9 -- 2460449.6) yields a marginal period detection at 0.08102(6) d, with a double-peaked profile and an amplitude of 0.07 mag. Due to the observation gap between the TESS sectors, we only observed this modulation for 8 cycles. Although we cannot fully exclude the possibility that this was genuine early superhumps due to the short coverage, this modulation resembles the pre-stage-A-superhump oscillations with a period close to the orbital period, observed in V844 Her and K2BS5 with a mass ratio larger than 0.1 \citep[][]{kat22v844her, boy24k2bs5}. These authors discussed that these pre-stage-A oscillations are not likely early superhumps because of the small number of observed cycles and the average epoch of humps being located on the smooth extension of stage-A superhumps. Combining all these features, V1676 Her is more likely to be a normal long-$P_\text{SH}$ system without the excitement of 2:1 resonance rather than a period bouncer.

\subsection{V748 Hya} \label{sec:v748hya}

V748 Hya (=ASAS J102522$-$1542.4) is one of the well-studied WZ Sge-type DN with the early and stage-B ordinary superhump periods of 0.06136(6) and 0.06337(2) d \citep[][]{Pdot}, suggested to be a borderline long-$P_{\text{orb}}$ WZ Sge-type DN.
Its previous outbursts were observed in January 2006 \citep[][]{van06asas0233asas1025, Pdot} and in February 2011 \citep[][]{Pdot3}. 
Its quiescence counterpart is Gaia EDR3 3749056714792756352 $G=19.27(1)$ mag at 281$^{+19}_{-20}$ pc, according to  Gaia EDR3 \citep[][]{gaiaedr3, Bai21GaiaEDR3distance}, corresponding to the absolute magnitude $M_G$ = 12.0(2) mag.
This system underwent an outburst in February 2025 (vsnet-alert 28048). The left panels of Fig. \ref{fig:oc-N5} present the observations of V748 Hya. The outburst rise was well captured by ATLAS and ZTF on BJD 2460729. The outburst peaked around BJD 2460730.9 at $g\simeq$11.3 mag, corresponding to the absolute magnitude $M_g$ = 4.1(2) mag, and entered the rapid decline before BJD 2460749.6. Thus, the amplitude and duration of this 2025 superoutburst are $\simeq$8.0 mag and 20 d, respectively.  After the rapid decline, a short and single rebrightening was observed on BJD 2460757 in ZTF, which was also observed in its 2006 superoutburst.
Moreover, because there is no record of other outbursts in ASAS-SN and ATLAS since 2015, the quiescence duration before this 2025 superoutburst is likely 14 years, triple that of the one between its 2006 and 2011 superoutbursts. This peak brightness is almost 1.0-mag brighter than that of the 2006 superoutburst (12.2 mag), although it could be just because of the lack of early coverage in 2006.

\subsubsection*{The 2025 superoutburst in TESS}

The 2025 superoutburst of V748 Hya was observed in the TESS sector 89. The rise timescale was 0.066(1) d mag$^{-1}$.
Based on the $O-C$ diagram, we determined the early, stage-A, and stage-B superhump phases as BJD 2460730.4 --  2460735.3, 2460736.7 -- 2460737.45, and 2460737.55 -- 2460744.7, respectively.
Through PDM analysis, the superhump periods are determined as 0.061861(9), 0.06423(5), and 0.063341(5)  d for early, stage-A, and stage-B superhumps, respectively. Early superhumps show a clear double-peaked profile with an amplitude of 0.04 mag. These superhump periods yield the mass ratio of V748 Hya as 0.103(3).
$P_{\text{dot}}$ during stage-B superhumps is determined as 12.5(2) $\times 10^{-5}$ cycle$^{-1}$.
The $O-C$ diagram suggests the existence of stage-C superhumps at $E \geq 235$, although the short duration prevents us from measuring their period.
These values of ordinary superhumps all agree with those obtained in its previous outburst from the ground telescopes within 3 sigma \citep[$P_\text{stage-A~SH},~P_\text{stage-B~SH},~\text{and}~P_\text{dot}$ are 0.0641(1) d, 0.06337(2) d, and 10.9(6) $\times 10^{-5}$ cycle$^{-1}$; ][]{Pdot}. The durations of the stage-B superhump stage are similar to each other ($\simeq$215 cycles in both superoutbursts). The early superhump period is slightly different (0.06136(6) d in 2006 and 0.061861(9) d in 2025), although \citet{Pdot} claimed that the baseline during the 2006 superoutburst may not be long enough. Thus, the true early superhump period is likely 0.061861(9) d.

\subsection{ASASSN-25ci} \label{sec:a25ci}

ASASSN-25ci was discovered by ASAS-SN in June 2025. The probable quiescence counterpart is GSC 2210321224608 \citep[$B = $21.04;][]{GSC2}. The outburst peaked around BJD 2460832.0 at $g=13.51$ mag and entered the rapid decline on BJD 2460852.1, giving the outburst duration as 20 d and the outburst amplitude as 7.5 mag (right panels of Fig. \ref{fig:oc-N5}). There are no associated Gaia EDR3 objects. ASASSN-25ci is classified as a WZ Sge-type DN candidate based on the late appearance of ordinary superhumps (vsnet-alert 28081). The reported superhump period is 0.0535(1) d.

\subsubsection*{The 2025 superoutburst in TESS}

TESS Sector 93 covered this entire superoutburst from before the outburst rise to after the end of the rapid decline, which is the best coverage among our samples. The rise timescale is 0.317(3) d mag$^{-1}$. 
We detect superhumps throughout the superoutburst. Judging from the $O-C$ diagram and superhump profile, we determined the early, ordinary stage-A, and stage-B superhump phases as BJD 2460831.5 -- 2460839.0, 2460840.9 -- 2460841.8, and 2460842.1 -- 2460849.9, respectively. The obtained periods are 0.052948(8), 0.05431(9), and 0.053656(9) d for the early, ordinary stage-A, and stage-B superhumps. The mean early superhump amplitude is 0.06 mag. The $P_{\text{dot}}$ in stage-B superhumps is 4.8(9) $\times 10^{-5}$ cycle$^{-1}$. Its mass ratio is estimated as 0.067(5) using the early and stage-A superhump periods.

\section{Discussion}
\label{sec:discussion}

\subsection{WZ Sge-type nature of our samples}

\begin{figure}
	\includegraphics[width=0.95\linewidth]{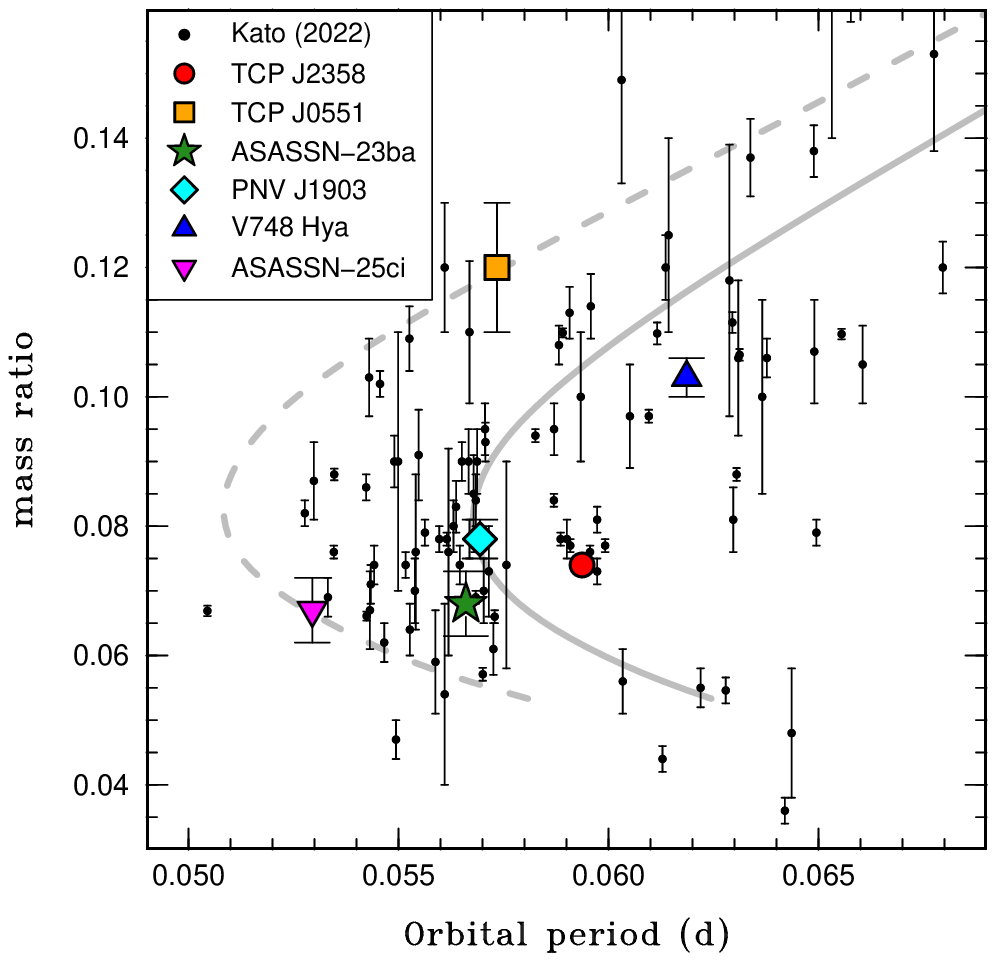}
    \caption{Orbital periods vs mass ratios of CVs around the period minimum. The black dots represent the samples in \citet{kat22updatedSHAmethod}. 
            The filled circle, square, star, diamond, upward triangle, and downward triangle show those of TCP J2358$+$55, TCP J0551$+$65, V1676 Her, PNV J1903$-$31, V748 Hya, and ASASSN-25ci, respectively. The solid and dashed lines represent the standard and optimal evolutionary path of CVs in \citet{kni11CVdonor}.}
    \label{fig:evol}
\end{figure}

We have analysed the TESS and ground-based time-resolved photometric observations of nine WZ Sge-type dwarf novae and candidates. Early superhumps are detected in five systems (TCP J2358$+$55; 0.059371(3) d, ASASSN-23ba;  0.056605(4) d, PNV J1903$-$31; 0.05694(1) d, V748 Hya; 0.061861(9) d, and ASASSN-25ci; 0.052948(8) d) using the TESS observations, confirming them as a WZ Sge-type DN.
TESS observations newly recognised early superhumps in PNV J1903$-$31 and ASASSN-25ci. The reason for missing their early superhumps in ground-based observations is a lack of enough coverage during the early phase of their superoutbursts. PNV J1903$-$31 showed relatively low-amplitude (0.02 mag) early superhumps, which are around the typical detection limit with ground observations. These points strengthen the importance of continuous observations such as TESS, even if the target is bright enough for small telescopes to detect early superhumps.

We detect the superhumps during the rebrightening phase of ASASSN-20mq. Given the long waiting time before the appearance of ordinary superhumps, its overall characteristics in outburst agree with its classification as a WZ Sge-type DN. The 2022 superoutburst of TCP J0551$+$65, which was unambiguously classified as a WZ Sge-type DN in its 2019 superoutburst, was observed by TESS. These observations detected ordinary superhumps immediately at the outburst maximum, confirming that this 2022 superoutburst lacked an early superhump phase. The outburst interval after the superoutburst with early superhumps was $\simeq$3.7 years, lying between AL Com and V627 Peg.

On the other hand, the WZ Sge-type nature in MO Psc and V1676 Her is questionable. The superhump period of MO Psc has been newly established as 0.06025(2) d with TESS, rather than 0.05161 d in \citet{shu21mopsc}. Together with its short ($\simeq$15 d) outburst duration, we suspect that MO Psc might be an SU UMa-type DN without the excitation of 2:1 resonance, although multiple rebrightening outbursts are more commonly observed in WZ Sge-type DNe. Future observations are vital to establish its classification. We confirmed the stage-B superhump period of V1676 Her in \citet{Pdot5}, and newly established its $P_\text{dot}$ as $-$16(1)$\times 10^{-5}$ cycle$^{-1}$. On the earliest days of the outburst, V1676 Her showed double-peaked oscillations with a period of 0.08102(6) d, shorter than that of stage-B superhumps. These overall features are similar to the superoutbursts in V844 Her and K2BS5 with large mass ratios, both studied with TESS and Kepler \citep[][]{kat22v844her, boy24k2bs5}. Thus, V1676 Her is not likely a period bouncer, in contrast to the speculations in \citet{Pdot5}.

We have determined the mass ratio of six samples using the early and ordinary stage-A superhump periods; five samples with early superhump detection by TESS and TCP J0551$+$65. Fig. \ref{fig:evol} presents the relation between the orbital periods and mass ratios of short-$P_{\text{orb}}$ CVs from \citet{kat22updatedSHAmethod} and our six samples. Among our samples, ASASSN-23ba, ASASSN-25ci, TCP J2358$+$55, and PNV J1903$-$31 have the mass ratios near or even below the period minimum within the error range \citep[$ q \simeq 0.08$; ][]{kni11CVdonor}. On the other hand, TCP J0551$+$65 and V748 Hya have $q \geq 0.1$, relatively large for a WZ Sge-type DNe. 
The mass ratio can also be estimated from the $P_{\text{dot}}$ during the stage-B superhump phase \citep[][]{Pdot, kat22updatedSHAmethod}. The obtained values of TCP J2358$+$55, V748 Hya, and ASASSN-25ci are 0.07(1), 0.11(2), and 0.08(2), all consistent with those estimated based on the superhump period excess of the stage-A superhumps. The large error in ASASSN-23ba does not give a meaningful constraint. Although the orbital (or early superhump) period of ASASSN-20mq is not available, given its stage-A superhump period (0.0559(1) d) and mass ratio (0.09(3)) estimated from $P_{\text{dot}}$, the system is presumably located slightly above the period minimum, which agrees with other WZ Sge-type DNe with the type-A rebrightening episode \citep[][]{kat22updatedSHAmethod}. The negative $P_\text{dot}$, large superhump amplitude, and long $P_\text{SH}$ of V1676 Her agree with a larger mass ratio than those of WZ Sge-type DNe.

\subsection{Superhump characteristics}
\label{sec:OCdiagram}

\begin{figure*}
	\includegraphics[width=\linewidth]{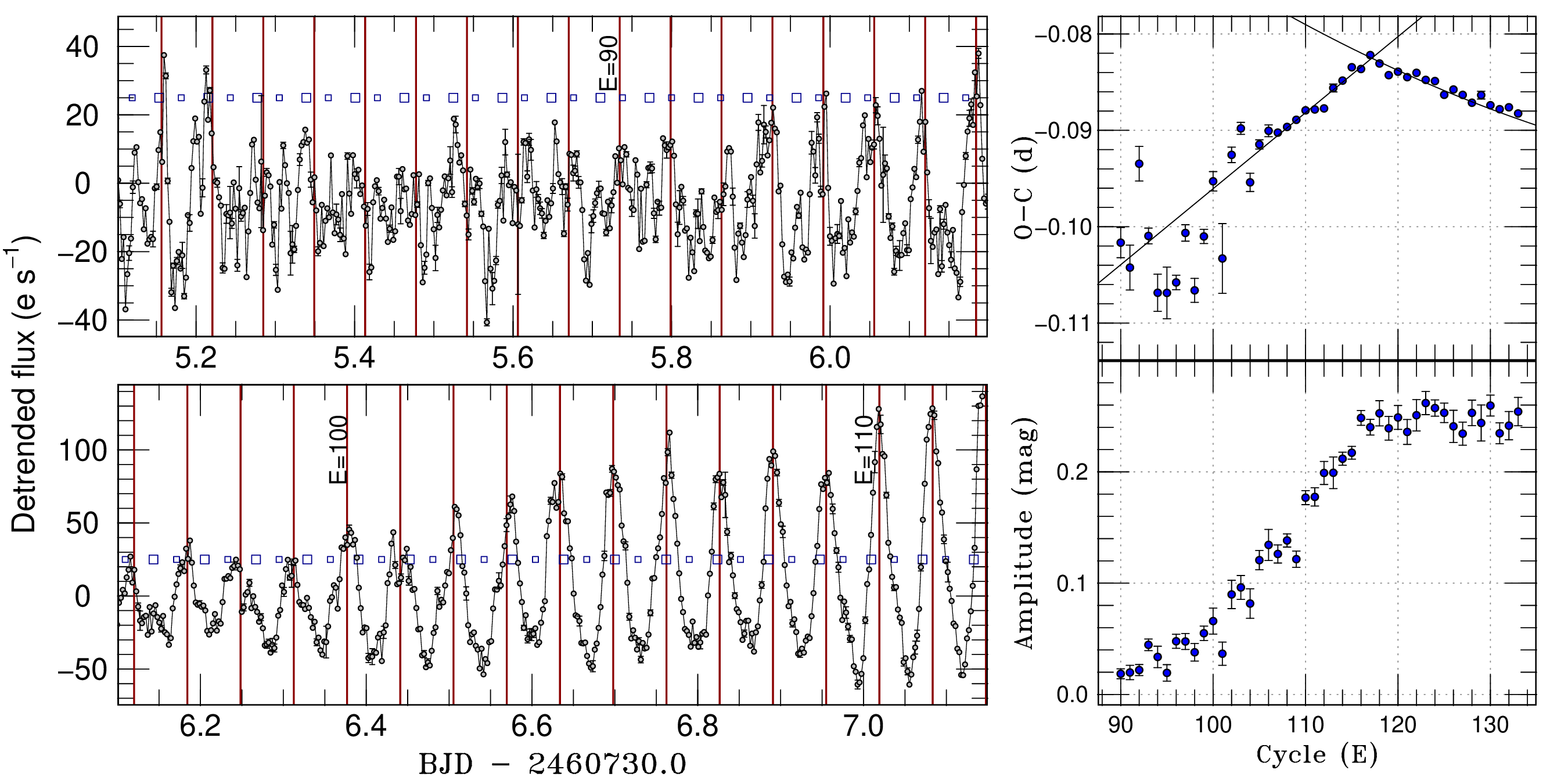}
    \caption{
        Left; zoomed light curve of V748 Hya around the early--ordinary superhump transition. The vertical lines present the epochs of stage-A superhump maxima based on the $O-C$ diagram. The large and small open squares indicate the epochs of the primary and secondary maxima of early superhumps.
        Right; the zoomed $O-C$ during the ordinary superhumps (top) of V748 Hya. The solid lines indicate the periods of stage-A and stage-B superhumps. The bottom panel represents the evolution of superhump amplitude on a magnitude scale. 
        }
    \label{fig:ocparafit}
\end{figure*}

As mentioned above, the wide range of mass ratios in our samples, together with the continuous coverage by TESS, is suitable for testing the existing relations in superhumps and the TTI model itself.
For example, \citet{kat22updatedSHAmethod} gives an updated relation between the orbital period $P_\text{orb}$ and the fractional superhump excess $\epsilon$ ($=P_\text{SH} / P_\text{orb} - 1$) of the mean stage B superhump periods. By adopting $P_\text{early~SH} = P_\text{orb}$, $\epsilon$ of TCP J2358$+$55, TCP J0551$+$65, ASASSN-23ba, PNV J1903$-$31, V748 Hya, ASASSN-25ci are 0.0116(1), 0.015(1), 0.0079(5), 0.0134(2), 0.0239(2), and 0.0134(2), all agree with other known systems.
The waiting times before the appearance of ordinary superhumps in our samples are $\simeq$5.5, 9.0, and 12.0 d in V748 Hya, ASASSN-25ci, and ASASSN-23ba. It took $\simeq5.5$ d in ASASSN-24hd with a mass ratio of 0.098(4)  \citep[][]{tam25asassn24hd}. The durations of stage-A superhumps of V748 Hya, ASASSN-25ci, and ASASSN-23ba are $\simeq$20, 20, and 30 cycles. In the TTI model, these durations are expected to be longer in more evolved systems, as the tidal resonances grow faster in a system with a higher mass ratio \citep[][]{osa05DImodel, kat15wzsge}.
\citet{kat22WZSgecandle} formulated the relation between the absolute magnitude at the appearance of ordinary superhumps and the amplitude of early superhumps, based on the assumption that (1) the disk size at the appearance of ordinary superhumps is expected to be around the 3:1 resonance radius and (2) the amplitude of early superhumps is correlated with the inclination of a system. Among our samples, V748 Hya and TCP J2358$+$55 have the Gaia EDR3 distances (281$^{+19}_{-20}$  and 835$^{+628}_{-344}$ pc, respectively). Thus, their magnitudes at the appearance of ordinary superhumps (12.6(1) and 14.0(1) mag, respectively) correspond to the absolute magnitude of 5.4(2) and 4.4(1.2) mag.  Equation 3 in \citet{kat22WZSgecandle} yields their mean amplitude of early superhumps (0.04 and 0.07 mag) to the absolute magnitudes 5.7(4) and 6.0(4) mag. These numbers agree within 3$\sigma$, although TCP J2358$+$55 might be slightly brighter than other systems implemented in this relation.

As presented in \citet{tam25asassn24hd}, TESS has a great capability for superhump studies in WZ Sge-type DNe, especially at the superhump stage transitions. V748 Hya is the best sample in this paper, given its bright outburst and relatively large amplitude of early superhumps. The left panels of Fig. \ref{fig:ocparafit}  present the normalised light curve around the appearance of ordinary superhumps. In this zoomed light curve, the epochs of maxima in each cycle agree with the expected times of the early superhump maximum up to BJD 2460735.65. After this, the maxima consistent with the expected times of the stage-A superhump maximum appear on BJD 2460735.74 ($E=90$). Although WZ Sge showed a smooth development of ordinary superhump maxima from the primary maxima of early superhumps \citep[][]{Pdot}, the ordinary superhump maxima may grow from the secondary maxima of early superhumps in V748 Hya, or possibly they are physically unrelated. The amplitude of ordinary superhumps steadily grows after this epoch (right bottom panel of Fig. \ref{fig:ocparafit}). On the other hand, the individual maxima are still a superposition of early and ordinary superhumps, resulting in a large scatter in the $O-C$ diagram during $E=$90--104 (right top panel of Fig. \ref{fig:ocparafit}). 
The stage A--B superhump transition occurred around BJD 2460737.47 ($E=117$), within $\simeq$2 ($E=$116--118) superhump cycles or $\simeq$0.12 d. The superhump amplitude reaches its maximum on BJD 2460737.40 ($E=116$). These behaviours are consistent with those of ASASSN-24hd \citep[][]{tam25asassn24hd}.

\subsection{Outburst rise}
\label{sec:LCrise}

\begin{figure*}
	\includegraphics[width=\linewidth]{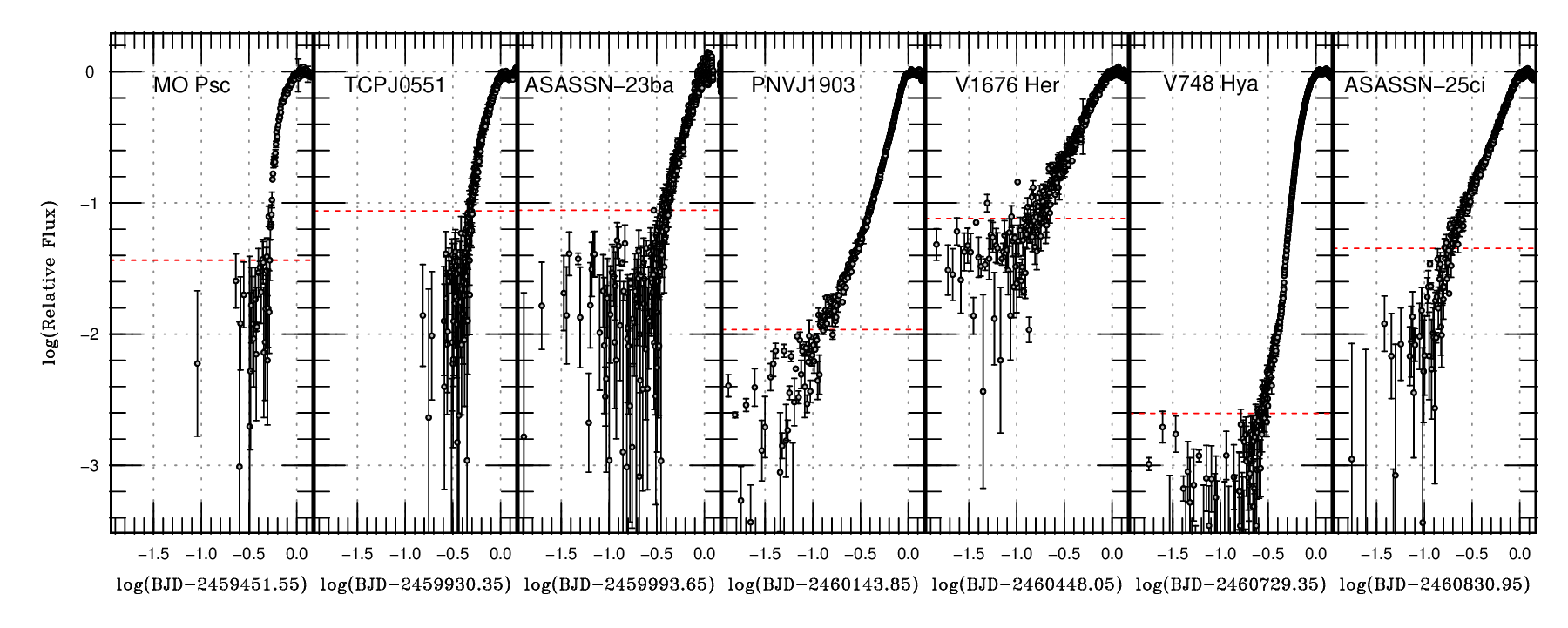}
    \caption{The zoomed light curves of MO Psc, TCP J0551$+$65, ASASSN-23ba, PNV J1903$-$31, V1676 Her, V748 Hya, and ASASSN-25ci around the outburst rise from left to right. The y-axis is normalised by the flux at the outburst maximum. The horizontal dashed lines present the 3$\sigma$ threshold of the differential flux before the outburst.}
    \label{fig:riselc}
\end{figure*}

\begin{figure*}
	\includegraphics[width=\linewidth]{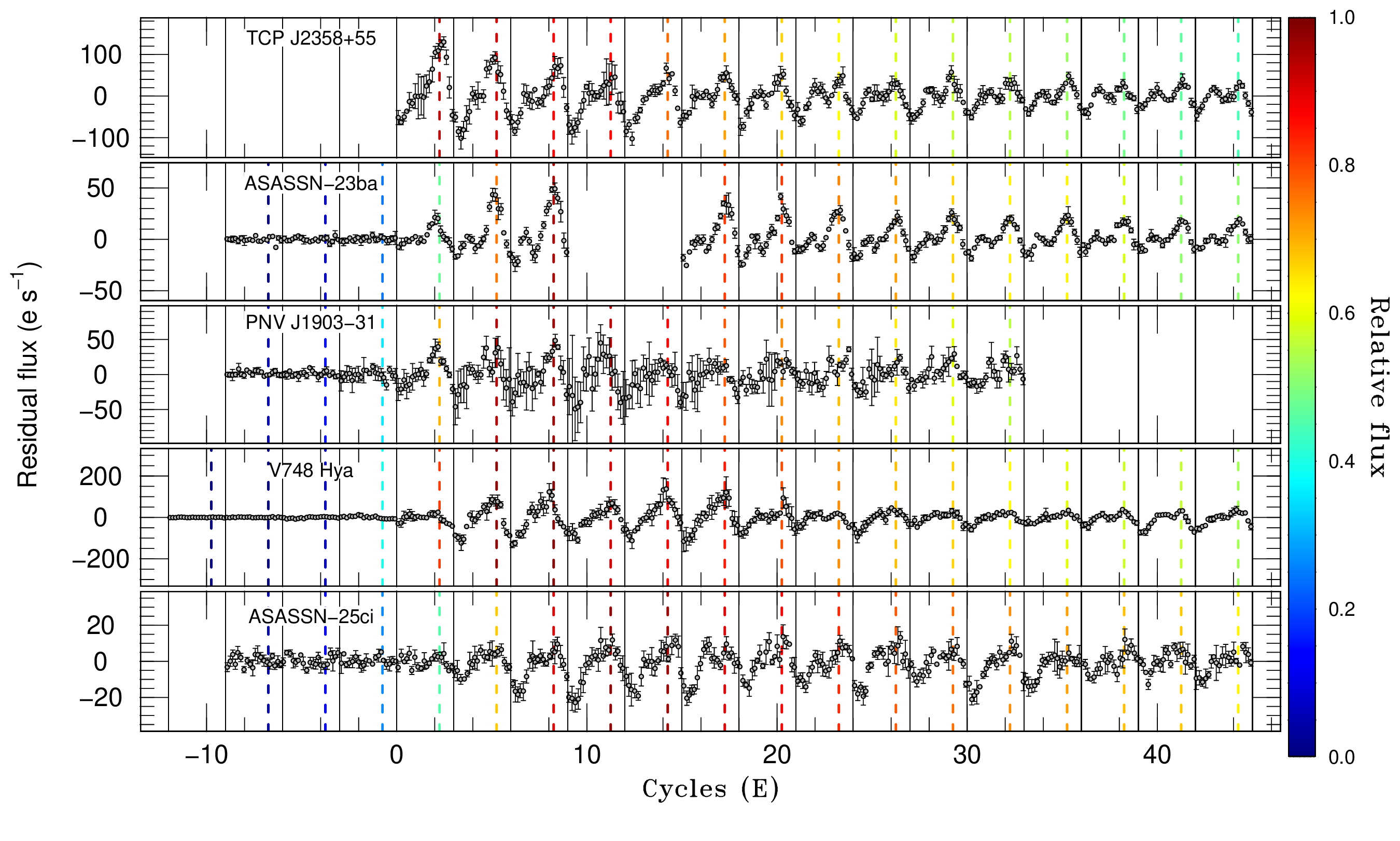}
    \caption{
            The phase-averaged early superhump profiles of TCP J2358$+$55, ASASSN-23ba, PNV J1903$-$31, V748 Hya, and ASASSN-25ci from top to bottom of the residual light curves in the TESS flux unit (e s$^{-1}$). Individual profiles are binned into each 3 early superhump cycles (e.g., $E = -3$--$-1$, $0$--$2$, and so on), and horizontally shifted to the respective $O-C$ cycles. The vertical lines present the phase of the early superhump maximum in the overall mean profile, and their colour indicates the median flux of the respective cycle compared to the outburst maximum flux.
        }
    \label{fig:risehump}
\end{figure*}

\begin{table}
	\centering
	\caption{Outburst rise timescales of our samples.}
	\label{tab:risetime}
	\begin{tabular}{rcccc} 
		\hline
		   & $\tau_{\rm rise}$ & $\alpha_1$ &  $\alpha_2$ &  Break\commenta \\
		   & [d mag$^{-1}$]  &  &  &  [hour] \\
		\hline
            MO Psc    & 0.125(3) & 3.8(1) & --- & --- \\
            TCP J0551$+$65  & 0.070(4) & 6.2(3) & --- & --- \\
            ASASSN-23ba    & 0.208(3) & 2.56(6) & --- & --- \\
            PNV J1903$-$31  & 0.169(1) & 2.03(8) & 2.89(1) & $-$14.0(2) \\
            V1676 Her    & 0.280(4) & 1.49(2) & --- & --- \\
            V748 Hya    & 0.066(1) & 3.6(3) & 8.03(7) & $-$13.85(8) \\
            ASASSN-25ci    & 0.317(3) & 1.87(1) & --- & --- \\
            \hline
            \multicolumn{5}{l}{\commenta The break time of the broken-powerlaw fitting}\\
            \multicolumn{5}{l}{respective to the epoch of the outburst maximum.}\\
	\end{tabular}
\end{table}

By its nature, WZ Sge-type DNe are considered to always trigger an outside-in outburst both in the thermal-tidal instability and mass-transfer burst models \citep[e.g.][]{ham97wzsgemodel, mey98wzsge}. The previously observed rise timescales of WZ Sge-type DNe are indeed fast; from $\simeq$0.1 \citep[V455 And and GW Lib; ][]{mae09v455andproc, vic11gwlib} to $\simeq$0.3 d mag$^{-1}$ (MASTER OT J030227.28$+$191754.5 in \citet{tam24j0302} and 4 samples in \citet{otu16DNstats}).
It must be noted, however, that the results in \citet{otu16DNstats} should be treated as an upper limit because their samples rely on the observations from the CRTS with 1-d cadences. 
Moreover, \citet{rid19j165350} reported the broken-powerlaw shape (indices of 0.95(9) and 4.82(7)) of the outburst rise in a WZ Sge-type DN KSN:BS-C11a. They reported that this trend was also observed in two other SU UMa-type DNe. However, it is technically very difficult for ground telescopes to capture and track the rise of an unpredictable outburst in WZ Sge-type DNe, and hence, the available samples are still limited.

Our TESS observations add seven samples with the coverage of the outburst rise: ASASSN-23ba, V1676 Her, ASASSN-25ci, V748 Hya, PNV J1903$-$31, MO Psc, and TCP J0551$+$65, increasing the number of samples by a factor of two. It must be noted that MO Psc and V1676 Her may not be WZ Sge-type DNe, and the 2022 superoutburst of TCP J0551$+$65 lacked the early superhump phase, though. 
Fig. \ref{fig:riselc} presents the light curves of these objects around the outburst rise in the flux scale normalised by their outburst peak flux. 
First, to provide a comparison with other WZ Sge-type DNe with ground observations, we measured the rise timescale (d mag$^{-1}$) by linearly fitting the light curve on a magnitude scale. 
We also fitted the light curve with the powerlaw function in the flux scale (e s$^{-1}$) following \citet{rid19j165350}. For V748 Hya and PNV J1903$-$31, we further fitted the light curve with a broken powerlaw function. Table \ref{tab:risetime} summarises the obtained rise timescales ($\tau_{\text{rise}}$) and powerlaw indices ($\alpha_1$ and $\alpha_2$) of our samples. V748 Hya shows a short rise timescale comparable to V455 And and GW Lib. The light curve shows a clear broken-powerlaw shape, characterised by the indices of 3.6(3) and 8.03(7), broken at 13.85(8) hours before the outburst maximum and around $\simeq1.3\%$ of the maximum flux. These features generally agree with KSN:BS-C11a \citep{rid19j165350}, although both the indices are larger in V748 Hya. The rise timescale of TCP J0551$+$65 is comparable to V748 Hya, but there is no break recorded in TESS.
PNV J1903$-$31, ASASSN-23ba, and ASASSN-25ci showed rise timescales within the range of the previous samples. Their powerlaw index is even smaller than that of V748 Hya before the break. PNV J1903$-$31 shows a broken-powerlaw type of the outburst rise; however, the change of the indices is only 40\% compared to a factor of $\simeq$2.2 and 5.0 in V748 Hya and KSN:BS-C11a, respectively. The break timing  (14.0(2) hours before the outburst maximum)  is similar to V748 Hya and at $\simeq$10\% of the peak flux.
The rise timescale of MO Psc and V1676 Her lies between these fast and slow populations.

Thus, our observations confirm V748 Hya as the second example of a broken-powerlaw rise in WZ Sge-type DNe, and possibly PNV J1903$-$31 as another example. Although the origin of such a broken-powerlaw rise is unclear, a similar broken-shape rise is presented in a simulated light curve of \citet{jor24ttisimulation}. In their work, the initial slow rise corresponds to the phase between the onset of an outburst at a specific radius and the start of the propagation of the heating wave across the entire disk. The travelling heating wave sets a faster rise.
It must be noted that the limiting flux of ASASSN-23ba, V1676 Her, ASASSN-25ci, TCP J0551$+$65, and MO Psc is well above the flux level of the break in V748 Hya and KSN:BS-C11a; 1\% of the peak flux or $\simeq$5 mag fainter than peak. 
The longest rise timescale of ASASSN-25ci among our samples is still much shorter than those found in the inside-out outbursts of SS Cyg \citep[][]{can98sscyg}. Moreover, since the rise timescales of outside-in outbursts of SS Cyg show a 1$\sigma$ range of 0.14 d mag$^{-1}$, the difference between V748 Hya and ASASSN-25ci is still within 2$\sigma$ of this.  Hence, all observed samples in this paper agree with an outside-in type of outbursts.

Fig. \ref{fig:risehump} presents the early superhump profiles of TCP J2358$+$55, ASASSN-23ba, PNV J1903$-$31, V748 Hya, and ASASSN-25ci from the outburst rise, binned in each 3 early superhump cycles. This clearly demonstrates the fast growth of the early superhump amplitude within $\sim$10 cycles as the system reaches the outburst maximum. In TCP J2358$+$55, PNV J1903$-$31, and V748 Hya, the secondary maxima of early superhumps are less prominent at the outburst maximum, and this becomes more evident in $E\geq15$. On the other hand, ASASSN-23ba shows clear secondary maxima already in $E\simeq$3--5 before the outburst maximum. The typical brightness of $E=0$ is 40--80\% of the peak flux (i.e., $\leq$1.0 mag fainter than the outburst maximum) except TCP J2358$+$55, which is not covered on the outburst rise. This is exactly the same behaviour in WZ Sge and V455 And \citep[][]{kat15wzsge}.

As mentioned in Section \ref{sec:intro}, the enhanced mass transfer model naturally predicts an enhanced emission from the hot spot. In eclipsing systems, early superhump maxima are located around the orbital phase $\phi\simeq0.6$. If there is any orbital hump originating from the hot spot in an extended and hot accretion disk (i.e., novalike stars), it should peak at $\phi\simeq0.9$ \citep[][]{rut92oycar}. 
We do not see this in any of our samples. Indeed, this is the phase where the primary minium of early superhumps is located (Fig. \ref{fig:pdm1}). 
We here constrain the orbital hump amplitude before the appearance of early superhumps in V748 Hya. 
The root-mean-square (RMS) of the residual flux before the appearance of early superhumps ($E=-15$--$0$) is consistent with a typical flux error at $\leq 3.0$ e s$^{-1}$. The peak brightness of V748 Hya was $o=$11.440(1) mag in ATLAS and $\simeq$3800 e s$^{-1}$ in TESS. Given its distance at 281$^{+19}_{-20}$ pc and assuming a blackbody spectral energy distribution of 12000 K, which is the typical colour temperature of WZ Sge-type DNe in outburst \citep[e.g.][]{mat09v455and, shu21aylac}, $\simeq3.0$ e s$^{-1}$ corresponds to emission from a 12000-K blackbody with a bolometric luminosity of $\simeq 1.0(2) \times 10^{31}$ erg s$^{-1}$, where the error originates from the uncertainty in distance. 
This can be compared with the bolometric luminosity of a hotspot, which is also known to be around 12000 K \citep[e.g.][]{woo86zcha, mar88ippeg}. \citet{sma02ADstructure} formulates the bolometric luminosity of a hotspot $L_\text{sp}$ as equation \ref{eq:hs} 

\begin{equation}
    \label{eq:hs}
    L_\text{sp} = \frac{1}{2} \Delta v^2 \left( \frac{2 \pi A}{P_\text{orb}} \right) \dot{M_\text{tr}}
\end{equation}

where $\dot{M_\text{tr}}$, $A$, and $\Delta v^2$ are the mass transfer rate from the secondary star, the binary separation, and a dimensionless quantity that depends on the fraction of disk radius over the Roche-lobe radius and weakly on the mass ratio. We here adopt the binary parameters of V748 Hya ($P_\text{orb}$ = 0.061861 d and $q=0.103$) and assume a $0.80M_\odot$ WD, which is typical of CVs below the period gap, including WZ Sge-type DNe \citep[][]{pal22WDinCVs}.
With its small mass ratio, the 2:1 resonance radius is almost equal to the Roche-lobe radius; hence, we assumed that the disk radius reaches the Roche-lobe radius beyond the tidal truncation radius. A smaller disk on the early rise of an outburst results in a more luminous hot spot. These points yield $A=4.4\times10^{10} \text{~cm,~}\Delta v^2=0.743$, and equation \ref{eq:hs1};

\begin{equation}
    \label{eq:hs1}
    L_\text{sp} \simeq 1.0 \times \left( \frac{\dot{M_\text{tr}}}{10^{15}\text{g~s}^{-1}} \right) \times10^{30}~\text{erg s}^{-1}
\end{equation}

The 0.04-mag early superhump amplitude of V748 Hya implies a relatively high inclination \citep[$i>60^\circ$ according to][]{kat22WZSgecandle}. \citet{neu23bwscl} discussed that one peak of the double-peaked orbital profile of BW Scl ($i=64.3(3.6)^\circ$) in quiescence is due to the hot spot shining through an optically-thin disc. This means, in turn, that the radially extended and optically-thick outer disk in the outburst rise would occult a hot spot at some orbital phases, and hence one expects to observe an orbital hump with an amplitude close to the hot spot luminosity. These discussions and equation \ref{eq:hs1} yield that the non-detection of orbital humps below the error level constrains the mass transfer rate to be lower than $\simeq 1\times10^{16}$ g s$^{-1}$ on the outburst rise. This is larger than a mass transfer rate normally found in quiescence of WZ Sge-type DNe \citep[an order of $\sim10^{15}$ g s$^{-1}$; e.g.][]{sma93wzsge, ama21ezlyn, neu23bwscl}. However, the implemented increase of the mass transfer rate in the enhanced mass transfer model \citep[$10^{19}$ ($10^{18}$) g s$^{-1}$ to transfer $10^{24}$ g in 1 (10) days; ][]{sma93wzsge, ham97wzsgemodel, mey98wzsge} is well above this constraint. Thus, our result on V748 Hya safely rules out a huge (factor of $\geq10$) enhancement of the mass transfer rate at the outburst rise demanded in the mass transfer burst model.

\section{Conclusions}
\label{sec:summary}

We report the TESS and ground-based time-resolved photometric observations of nine WZ Sge-type DNe and candidates. Our key findings are summarised as follows; 

\begin{itemize}
    \item 
        We detect early superhumps, which are the ambiguous feature of a WZ Sge-type DN, from TESS in five systems (TCP J2358$+$55; 0.059371(3) d, ASASSN-23ba;  0.056605(4) d, PNV J1903$-$31; 0.05694(1) d, V748 Hya; 0.061861(9) d, and ASASSN-25ci; 0.052948(8) d), among which PNV J1903$-$31 and ASASSN-25ci are the first reported cases. In addition, the overall superoutburst properties and long-lived rebrightening outburst of ASASSN-20mq agree with its WZ Sge-type classification. A confirmed WZ Sge-type DN TCP J0551$+$65 showed a superoutburst lacking the early superhump phase, joining the minor population in WZ Sge-type DNe that show superoutbursts both with and without early superhumps.  The mass ratios of TCP J2358$+$55, TCP J0551$+$65, ASASSN-23ba, PNV J1903$-$31, V748 Hya, and ASASSN-25ci are measured as 0.074(1), 0.12(1), 0.068(5), 0.078(3), 0.103(3), and 0.067(5), respectively.

    \item 
        We find the superhump period of MO Psc as 0.06025(2) d, significantly longer than the previously reported value. Together with its short outburst duration (15.0 d), MO Psc might not be a WZ Sge-type DN, although it underwent multiple rebrightening outbursts. Although V1676 Her has been suspected to be a period bouncer, the obtained negative $P_\text{dot}$ during the stage-B superhumps suggests that V1676 Her is rather a normal long-$P_\text{orb}$ SU UMa-type DN.

    \item 
        In V748 Hya, the maxima of ordinary superhumps possibly grow from the secondary maxima of early superhumps, or they are physically unrelated, unlike the case in WZ Sge. The early superhump maxima affected the determination of superhump maxima during this co-existing phase, resulting in a large scatter in the $O-C$ diagram. The stage transition between stage-A and stage-B superhumps occurred within $\simeq$2 superhump cycles, confirming the result in ASASSN-24hd also obtained with TESS. The superhump amplitude reaches its maximum value almost at the same cycle as the stage A--B superhump state transition.
    
    \item
        We find the broken-powerlaw rise in the TESS light curves of V748 Hya and PNV J1903$-$31, joining KSN:BS-C11a observed by Kepler. We also measured the rise timescales of seven of our samples, which increases the number of samples by a factor of 2. All of their rise timescales are consistent with an outside-in type outburst, as expected in WZ Sge-type DNe. 

    \item 
        The early superhumps appeared when the system is at 40--80\% of the outburst peak flux. Before and after this epoch, we do not find any orbital humps associated with a hotspot. These non-detections in V748 Hya yield the hotspot luminosity $L_\text{sp} \leq 1\times10^{31}$ erg s$^{-1}$ and corresponding mass transfer rate $\dot{M}_\text{tr} \leq 1\times10^{16}$ g s$^{-1}$ on the early outburst rise. Thus, an enhancement of the mass transfer rate, even if it occurs, must be smaller than a factor of $\simeq$10 from the quiescence, which disagrees with the expected enhanced mass-transfer rate in the mass-transfer burst model.

\end{itemize}

\section*{Acknowledgements}


We acknowledge amateur and professional astronomers around the world who have shared data on variable stars and transients with the VSNET collaboration. This paper uses observations made from the South African Astronomical Observatory (SAAO). 
This work was supported by the Slovak Research and Development Agency under the contract No. APVV-20-0148.
We are grateful for financial support from grants APVV-20-0148, VEGA 2/0030/21, and VEGA 2/0003/25.

This work has made use of data from the Asteroid Terrestrial-impact Last Alert System (ATLAS) project. The ATLAS project is primarily funded to search for near earth asteroids through NASA grants NN12AR55G, 80NSSC18K0284, and 80NSSC18K1575; byproducts of the NEO search include images and catalogs from the survey area. This work was partially funded by Kepler/K2 grant J1944/80NSSC19K0112 and HST GO-15889, and STFC grants ST/T000198/1 and ST/S006109/1. The ATLAS science products have been made possible through the contributions of the University of Hawaii Institute for Astronomy, the Queen’s University Belfast, the Space Telescope Science Institute, the South African Astronomical Observatory, and The Millennium Institute of Astrophysics (MAS), Chile.

\section*{Data Availability}


The TESS FFIs are publicly available at the MAST and  TESScut (\url{https://mast.stsci.edu/tesscut/}). \texttt{TESSreduce} package \citep{rid21tessreduce} is a public code accessible at \url{https://github.com/CheerfulUser/TESSreduce/tree/master}. Most of the ground-based observations are publicly available at the American Association of Variable Star Observers (AAVSO) International Database (\url{https://www.aavso.org/}) and the Variable Star Observers League in Japan (VSOLJ; \url{https://vsolj.cetus-net.org/}). Other data is available upon request to individual observers.



\bibliographystyle{mnras}
\bibliography{cvs} 




\appendix





\bsp	
\label{lastpage}
\end{document}